\documentclass[usenatbib]{mnras}
\usepackage{graphicx}
\usepackage{amsmath}
\usepackage{breqn}
\usepackage{amssymb}
\usepackage{hyperref}
\usepackage{xcolor}

\title[Impact of gas spin and LW flux on BH seeding]{Impact of gas spin and Lyman-Werner flux on black hole seed formation in cosmological simulations: implications for direct collapse}
\author[Bhowmick et al.]{Aklant K. Bhowmick$^{1}$,
Laura Blecha$^{1}$,
Paul Torrey$^{2}$,
Luke Zoltan Kelley$^{3}$, \newauthor
Mark Vogelsberger$^{4}$,
Dylan Nelson$^{5}$,
Rainer Weinberger$^{6}$,
Lars Hernquist$^{6}$
\\
$^{1}$Dept.~of Physics, University of Florida, Gainesville, FL 32611, USA\\
$^{2}$Dept.~of Astronomy, University of Florida, Gainesville, FL 32611, USA\\
$^{3}$Dept.~of Physics and Astronomy, Northwestern University, Evanston, IL 60208, United States\\
$^{4}$Dept.~of Physics, Kavli Institute for Astrophysics and Space Research, Massachusetts Institute of Technology,
Cambridge, MA 02139, USA \\
$^{5}$Universit\"{a}t Heidelberg, Zentrum f\"{u}r Astronomie, Institut f\"{u}r theoretische Astrophysik, Albert-Ueberle-Str. 2, 69120 Heidelberg, Germany\\
$^{6}$Harvard-Smithsonian Center for Astrophysics, 60 Garden Street, Cambridge, MA 02138, USA\\
}
\begin{document}
\maketitle
\begin{abstract}
Direct collapse black holes~(BH) are promising candidates for producing massive $z\gtrsim 6$ quasars, but their formation requires fine-tuned conditions. In this work, we use cosmological zoom simulations to study systematically the impact of requiring: 1) low gas angular momentum~(spin), and 2) a minimum incident Lyman-Werner~(LW) flux in order to form BH seeds. We probe the formation of seeds~(with initial masses of $M_{\rm seed} \sim 10^4$ - $10^6 M_{\odot}/h)$ in halos with a total mass $> 3000\times M_{\mathrm{seed}}$ and a dense, metal poor gas mass $> 5\times M_{\mathrm{seed}}$. Within this framework, we find that the seed-forming halos have a prior history of star formation and metal enrichment, but they also contain pockets of dense, metal poor gas. When seeding is further restricted to halos with low gas spins, the number of seeds formed is suppressed by factors of $\sim6$ compared to the baseline model, regardless of the seed mass. Seed formation is much more strongly impacted if the dense, metal poor gas is required to have a critical LW flux~($J_{\mathrm{crit}}$). Even for $J_{\mathrm{crit}}$ values as low as $50~J_{21}$, no $8\times10^{5}~M_{\odot}/h$ seeds are formed. While lower mass~($1.25\times10^{4},1\times10^{5}~M_{\odot}/h$) seeds do form, they are strongly suppressed~(by factors of $\sim10-100$) compared to the baseline model at gas mass resolutions of $\sim10^4~M_{\odot}/h$~(with even stronger suppression at higher resolutions). As a result, BH merger rates are also similarly suppressed. Since early BH growth is dominated by mergers in our models, none of the seeds are able to grow to the supermassive regime~($\gtrsim10^6~M_{\odot}/h$) by $z=7$. Our results hint that producing the bulk of the $z\gtrsim6$ supermassive BH population may require alternate seeding scenarios that do not depend on the LW flux, early BH growth dominated by rapid or super-Eddington accretion, or a combination of these possibilities.
\end{abstract}



\section{Introduction}
\label{Introduction}

Supermassive black holes~(SMBHs) are now believed to be central components of galaxy formation and evolution. Almost every massive galaxy in the local Universe harbors a SMBH~\citep{1992ApJ...393..559K,1994ApJ...435L..35H,1995Natur.373..127M}. Evidence for SMBHs is also seen at higher redshifts~($z\gtrsim1$), where they are primarily observed as active galactic nuclei~(AGN). The most luminous AGN~(a.k.a quasars) have now been observed to redshifts of $z\sim7.5$~\citep{2001AJ....122.2833F,2011Natur.474..616M,2015Natur.518..512W,2018Natur.553..473B}. However, these quasars likely represent a very tiny and highly biased portion of the underlying SMBH population at $z\gtrsim7$; this population is going to be unveiled by upcoming facilities such as James Webb Space Telescope 
\citep[JWST;][]{2006SSRv..123..485G}, the Nancy Graham Roman Space telescope 
(NGRST, formerly WFIRST; \citealt{2015arXiv150303757S}), the Lynx X-ray Observatory 
\citep{2018arXiv180909642T}, and the Laser Interferometer Space Antenna  
\citep[LISA;][]{2019arXiv190706482B}. The overall $z\gtrsim7$ SMBH population~(including the observed brightest quasars) potentially contains imprints of the earliest seeds of SMBHs, which is currently a major theoretical gap in contemporary galaxy formation models. 

A popular candidate for SMBH seeds is the remnant of a first generation~(Population III or Pop III) star~(e.g., \citealt{2001ApJ...551L..27M}; \citealt{Bromm_2002}; \citealt{2003ApJ...593..661V}; \citealt{10.1093/mnras/sty3298}; \citealt{Smith_2018}; see also review by \citealt{doi:10.1146/annurev-astro-120419-014455} and references therein). This is a very promising channel for potentially explaining a substantial fraction of SMBHs, largely because they almost certainly exist as an inevitable consequence of the collapse of such massive stars~($\sim10-1000~M_{
\odot}$). Seeds formed via this channel are predicted to have masses $\sim100~M_{\odot}$~\citep{2001ApJ...550..372F}. However, the inferred masses of the $z>7$ quasars~($\sim10^9~M_{\odot}/h$) pose a huge challenge to Pop III seeds, since they require sustained accretion of gas at the Eddington limit to grow by $\sim7$ orders of magnitude by $z\sim7$. Alternatively, a higher initial seed mass~($\sim10^4-10^6~M_{\odot}/h$) makes it substantially easier for BH seeds to grow to $\gtrsim 10^9~ M_{\odot}/h$ by $z\sim7$. For this reason, black holes formed from direct collapse of pristine gas~(a.k.a. ``direct collapse black holes" or ``DCBHs") 
have become popular candidates for $z>7$ quasar progenitors, particularly because this channel can potentially form very massive seeds between $\sim10^4-10^6~M_{\odot}/h$~\citep[e.g.,][]{2003ApJ...596...34B,2006MNRAS.370..289B,2014ApJ...795..137R,2016MNRAS.458..233L,2018MNRAS.476.3523L,Becerra18,2019Natur.566...85W,2020MNRAS.492.4917L}.

In order to form $\sim10^4-10^6~M_{\odot}/h$ DCBHs, gas needs to undergo a nearly isothermal collapse at temperatures $\gtrsim10^4$ K. Additionally, large inflow rates~($\gtrsim0.1~M_{\odot}\mathrm{yr}^{-1}$ at a few tens of $\mathrm{pc}$ scales sustained for $\sim10~\mathrm{Myr}$) are needed to form a massive compact object~\citep[e.g.,][]{2010MNRAS.402..673B,2012ApJ...756...93H,2013ApJ...778..178H,2013A&A...558A..59S,2020OJAp....3E...9R,2021A&A...652L...7H}. To sustain the gas at $\gtrsim10^4$ K and make it eventually collapse, metal-line cooling and molecular hydrogen cooling channels need to be suppressed until the host halo assembles at least enough mass to cross the virial temperature $T_{\mathrm{vir}}\sim10^{4}$ K (known as 
the atomic cooling threshold, corresponding to halo masses $\sim10^7~M_{\odot}$). Primordial gas is devoid of metals, so molecular hydrogen is the only agent that can cool below $\sim10^4$ K and fragment the gas into forming the first generation of Pop III stars. 
Once star formation begins, subsequent stellar evolution will pollute the gas with metals, and that region will longer be able to form DCBHs.

Molecular hydrogen formation can be suppressed if the gas is exposed to a sufficient amount of UV radiation in the Lyman Werner~(LW) band~($11.2-13.5~\mathrm{eV}$). The minimum amount of LW flux ($J_{\mathrm{crit}}$) required to suppress fragmentation depends crucially on the source radiation spectrum~\citep{2014MNRAS.443.1979L}, self sheilding of $H_2$~\citep{2011MNRAS.418..838W} and the modeling of gas chemistry~\citep{2015MNRAS.451.2082G}. At the mini-halo stage, this could be achieved with low values of $J_{\mathrm{crit}}\sim 10^{-4}-1~J_{21}$~(where $J_{21}=10^{-21}\mathrm{erg \: s^{-1} cm^{-2} Hz^{-1} sr^{-1}}$) that can be supplied by the mean LW background~\citep{2014MNRAS.445.1056V}. However, $J_{\mathrm{crit}}$ steeply increases with halo mass; by the time halos cross the atomic cooling threshold, $J_{\mathrm{crit}}$ values are high enough that $H_2$ can only be dissociated if there is a nearby star burst of young Population II~(Pop II) and Pop III stars. Estimated values of $J_{\mathrm{crit}}$ in this regime are typically $\gtrsim 1000~J_{21}$ 
based on radiation hydrodynamic simulations~\citep{2010MNRAS.402.1249S}  as well as one-zone chemistry models~\citep{2014MNRAS.445..544S,2017MNRAS.469.3329W}. That being said, some recent works have also found that dynamical heating in halos~(triggered by periods of rapid growth) can significantly contribute to the 
suppression of cooling, thereby further decreasing $J_{\mathrm{crit}}$ to $\sim3~J_{21}$~\citep{2019Natur.566...85W,2020OJAp....3E..15R}.

Another potential impediment for DCBH formation is the angular momentum~(spin) of the gas. In addition to the suppression of molecular hydrogen cooling, having low gas spin may also be necessary to achieve inflow rates $\gtrsim0.1~M_{\odot}\mathrm{yr}^{-1}$. In other words, halos with high gas spin can provide rotational support to the pre-galactic gas disc and prevent the gas from achieving such high inflow rates. 

While DCBHs are a promising alternative to alleviate the stringent growth timescales of lower mass~($\lesssim10^3~M_{\odot}$) seeds, the previous considerations make it clear that their formation requires a number of fairly restrictive 
conditions to be simultaneously satisfied. This raises 
a couple of broad questions: 1) How (un)common are DCBH-forming gas environments within a given large scale structure? 2) What portion of the observable SMBH population originates from DCBH seeds?

Several aspects of these questions have been investigated in previous works; a vast majority of them use semi-empirical models~\citep{2006MNRAS.371.1813L,2007MNRAS.377L..64L,2012MNRAS.422.2051N,2008MNRAS.391.1961D,2014MNRAS.442.2036D,2018MNRAS.481.3278R,2020MNRAS.491.4973D}. \cite{2006MNRAS.371.1813L} found that only $\sim5\%$ of dark matter halos~($\sim10^7~M_{\odot}$) have spins that are low enough to form DCBHs~($\sim10^5~M_{\odot}$). \cite{2008MNRAS.391.1961D} found that a very small fraction~($10^{-8}-10^{-6}$) of atomic cooling halos have a close star forming neighbor~($\lesssim10~\mathrm{kpc}$) that provides LW fluxes~($\gtrsim10^3~J_{21}$) necessary to prevent fragmentation. Additionally, a halo may also need to be ``synchronized" with a nearby star forming halo---i.e., both halos need to cross the atomic cooling threshold within a few $\sim5$ Myr apart from each other~\citep{2014MNRAS.445.1056V,2017NatAs...1E..75R,2021MNRAS.503.5046L}. Overall, these results indicate that DCBH formation sites may be rare. While they could potentially explain the rarest, brightest tip of the observed high-z SMBH population, accounting for the ``typical"~(lower masses and luminosities, yet to be observed) SMBHs at these redshifts but may be a lot more difficult.

Recently, cosmological hydrodynamic simulations \citep[see][for a recent review]{2020NatRP...2...42V} have been used to probe the large scale structure for DCBH formation sites~\citep{2016MNRAS.463..529H,2017MNRAS.470.1121T,2018ApJ...861...39D,2020MNRAS.492.4917L,2021MNRAS.502..700C}. While they are considerably more expensive than semi-empirical models, they have the unique advantage of being able to self-consistently track the dynamics of gas which is a crucial ingredient for governing seed formation. This makes them an ideal tool to systematically assess the importance of different seeding criteria on DCBH formation. For example, \cite{2016MNRAS.463..529H}~(hereafter H16) studied the impact of varying $J_{\mathrm{crit}}$ using simulations spanning a wide range of volumes and resolutions. \cite{2018ApJ...861...39D} performed a similar study, seeding DCBHs using only the local gas properties instead of those averaged over the entire host halo. In principle, the use of local gas properties is more physically consistent than halo averaged gas properties~(as done in \citealt{2021arXiv210508055B} where the seeding is based on the total mass and star forming, metal poor gas mass of a halo); this is particularly true for modeling seeding conditions such as high densities and low metallicities. However, the actual length scales that are referred to as ``local" are determined by the spatial resolution of the simulation. Achieving resolution convergence may be more challenging when the seeding is only based on properties of a single ``local" gas cell. Moreover, seeding conditions based on properties such as gas angular momentum inevitably require information beyond the local environment. Consequently, \cite{2018ApJ...861...39D} decided not to explore the impact of gas angular momentum on seed formation. At the same time, H16 also decided to focus only on the LW flux criterion. It is important to note that \cite{2018ApJ...861...39D} does not resolve mini-halos, which can form stars and get polluted with metals before crossing the atomic cooling threshold; this will tend to overestimate the number of potential DCBH forming halos. 

In this work, we use a suite of cosmological hydrodynamic zoom simulations to systematically characterize the impact of both Lyman Werner flux and gas angular momentum  based seeding criteria on the SMBH population at $z>7$. We specifically investigate how seeds of different birth masses grow in the presence of these seeding criteria. Because of the fine-tuned nature of DCBH seeding conditions, probing the rare conditions of metal-poor gas irradiated by LW fluxes $\gtrsim 1000~J_{21}$ would require simulating a large cosmological volume. We find that 
our zoom region is only able to probe $J_{\mathrm{crit}}$ values up to $100~J_{21}$. We are not able to directly probe much higher values~$\gtrsim1000~J_{21}$ that are more likely to represent actual DCBH formation; nevertheless, our results will still enable us to assess the feasibility of the DCBH channel to explain different parts of the underlying mass function of $z>7$ SMBHs. In a follow-up paper~(Bhowmick et al in prep), we are probing the formation of the brightest $z>6$ quasars in much more extreme regions where 
higher LW fluxes are expected.

We also note that, similar to  \cite{2018ApJ...861...39D}, we 
do not resolve star formation in mini-halos; as a result, we do not intend to probe the typical DCBH formation scenario wherein halos just crossing the atomic cooling threshold are exposed to supercritical LW fluxes. Instead, we will focus on possible formation of seeds in halos that have grown significantly~(by factors of $\gtrsim4-10$ depending on the seed mass) beyond the atomic cooling threshold by the time they are exposed to supercritical LW fluxes. A majority of these halos have already initiated star formation but they do not get instantaneously metal enriched; as a result, some halos may have pockets of metal poor gas that can potentially be exposed to supercritical fluxes from nearby star forming regions. As it turns out, our model ends up  producing seeds in these metal poor pockets of dense gas. Therefore, it is this scenario that is the focus of our present study. 

This work is also part of a larger effort~(started with \citealt{2021arXiv210508055B}) to build a family of gas based seeding prescriptions for the next generation of cosmological simulations. Our prescriptions are generally agnostic about which theoretical seed formation channels 
they may represent (e.g., Pop III, DCBH or something else), such that we can tune our parameters to emulate a specific model. To that end, we use zoom simulations to characterize the impact of various aspects of galaxy evolution on the formation of seeds and their subsequent growth. 

This paper is organized as follows. Section \ref{Methods} presents the basic methodology, which includes the simulation suite, and the implementation and gas angular momentum and LW flux based seeding.  Section \ref{Results} describes the results of our work, followed by our main conclusions in Section \ref{Summary and Conclusions}.


\section{Methods}
\label{Methods}

Our simulations were run using the \texttt{AREPO} code~\citep{2010MNRAS.401..791S,2011MNRAS.418.1392P,2016MNRAS.462.2603P,2020ApJS..248...32W} which solves for gravity coupled with magnetohydrodynamics~(MHD). 
The gravity sector involves an N-body solver using PM-Tree method~\citep{1986Natur.324..446B}. The MHD sector uses a quasi-Lagrangian description of the gas fluid with an unstructured grid constructed via a Voronoi tessellation of the domain. \texttt{AREPO} has been used to run a variety of cosmological simulations which include uniform volumes such as Illustris~\citep{2014Natur.509..177V,2014MNRAS.444.1518V,2014MNRAS.445..175G,2015A&C....13...12N,2015MNRAS.452..575S},  IllustrisTNG~\citep{2018MNRAS.475..648P,2018MNRAS.475..624N,2018MNRAS.480.5113M,2018MNRAS.477.1206N,2018MNRAS.475..676S,2019MNRAS.490.3196P,2019ComAC...6....2N,2019MNRAS.490.3234N}, and zoom volumes such as AURIGA~\citep{2017MNRAS.467..179G} and HESTIA~\citep{2020MNRAS.498.2968L}.

\texttt{AREPO} contains several distinct sets of galaxy formation models. As in \cite{2021arXiv210508055B}, we use the IllustrisTNG model~\citep{2017MNRAS.465.3291W, 2018MNRAS.473.4077P} in this work as our baseline~(except for the BH seed model). The key features of the IllustrisTNG model include: star formation in the dense interstellar medium, where stars form stochastically from gas cells ~(with an associated time scale of $2.2~\mathrm{Gyr}$) when their densities exceed a threshold of $\rho_{\mathrm{SF}}=0.13~\mathrm{cm}^{-3}$, and the ISM itself is modelled by an effective equation of state \citep{2003MNRAS.339..289S,2013MNRAS.436.3031V}. In doing so, our model assumes that each gas cell with density $>0.13 ~\mathrm{cm}^{-3}$ has an unresolved cold dense component that can form stars; in regions where the gas is pristine, this cold dense component is presumably formed via molecular~($H_2$) cooling that is otherwise not explicitly included in the model. 

The absence of an explicit modeling for $H_2$ cooling will artificially suppress star formation and metal enrichment in mini-halos~(with virial temperature $T_{\mathrm{vir}}<10^4~\mathrm{K}$). While we do not resolve minihalos below $\lesssim10^6~M_{\odot}/h$, this can also over-estimate the number of metal-poor halos crossing the atomic cooling threshold. As we shall see, with a minimum halo mass criterion and a LW-flux-based seeding criterion~($J_{\mathrm{crit}}\geq50~J_{21}$), our simulation ends up largely forming seeds in $10^8\lesssim M_h \lesssim 10^{10}~M_{\odot}/h$ halos; 
these halos have grown significantly since crossing the atomic cooling threshold. We again emphasize that the existence of metal-poor pockets of gas within these star-forming halos allows seeding to occur even after the onset of star formation in a given halo. Owing to this, and because our galaxy formation model is well calibrated for these halos, the lack of explicit $H_2$ cooling model should not have a serious impact on our results. 
Note also that the exclusion of an explicit $H_2$ cooling model is a feature of many large volume cosmological simulations~\citep[for e.g.][]{2014MNRAS.444.1518V,2015MNRAS.450.1349K,2015MNRAS.446..521S,2016MNRAS.463..529H,2018MNRAS.475..624N}, and coincides with the choice not to resolve the cold, dense phase of the interstellar medium. We plan to explore this issue further in future work.

Stellar evolution and metal enrichment assumes a \cite{2003PASP..115..763C} initial mass function for the underlying single stellar populations~(SSPs) represented by the star particles. Metal cooling is implemented in the presence of a spatially uniform and time dependent ultraviolet background~(UVB) radiation field~(including the self shielding of dense gas). A uniform seed magnetic field~($10^{-14}$ comoving Gauss) is added at an arbitrary orientation, and its subsequent evolution is governed by MHD.

The modelling of black hole seeding is discussed in detail in the following section. Here, we summarize the other aspects of the black hole modeling that have been kept the same as IllustrisTNG. Once seeded, BHs can grow either via gas accretion or mergers with other BHs. BH accretion follows the Eddington limited Bondi-Hoyle formula and is given by 
\begin{eqnarray}
\dot{M}_{\mathrm{BH}}=\mathrm{min}(\dot{M}_{\mathrm{Bondi}}, \dot{M}_{\mathrm{Edd}})\\
\dot{M}_{\mathrm{Bondi}}=\frac{4 \pi G^2 M_{\mathrm{BH}}^2 \rho}{c_s^3}\\
\dot{M}_{\mathrm{Edd}}=\frac{4\pi G M_{\mathrm{BH}} m_p}{\epsilon_r \sigma_T}c
\label{bondi_eqn}
\end{eqnarray}
where $G$ is the gravitational constant, $M_{\mathrm{BH}}$ is the mass of the BH, $\rho$ is the local gas density, $c_s$ is the local sound speed of the gas, $m_p$ is the mass of the proton, $\epsilon_r$ is the radiative efficiency and $\sigma_T$ is the Thompson scattering cross section. The resulting bolometric luminosity is given by 
\begin{equation}
    L=\epsilon_r \dot{M}_{\mathrm{BH}} c^2,
    \label{bolometric_lum_eqn}
\end{equation}
where $\epsilon_r=0.2$. Accreting BHs inject energy into the surrounding gas as Active Galactic Nuclei~(AGN) feedback; this is implemented in two modes. For Eddington ratios~($\eta \equiv \dot{M}_{\mathrm{bh}}/\dot{M}_{\mathrm{Edd}}$) higher than  a critical threshold of  $\eta_{\mathrm{crit}}=\mathrm{min}[0.002(M_{\mathrm{BH}}/10^8 M_{\odot})^2,0.1]$, thermal energy is injected into the neighboring gas at a rate given by $\epsilon_{f,\mathrm{high}} \epsilon_r \dot{M}_{\mathrm{BH}}c^2$, where $ ~\epsilon_{f,\mathrm{high}} \epsilon_r=0.02$; $\epsilon_{f,\mathrm{high}}$ is referred to as the ``high accretion state" coupling efficiency. For Eddington ratio values lower than the critical threshold, kinetic energy is injected into the gas surrounding the black hole, in a time pulsated fashion, as a directed `wind' oriented along a randomly chosen direction; the energy injection rate~($\dot{E}_{\mathrm{kin}}$) is given by 
\begin{eqnarray}
    \dot{E}_{\mathrm{kin}}=\epsilon_{f,\mathrm{kin}} \dot{M}_{\mathrm{BH}}c^2, \\    \epsilon_{f,\mathrm{kin}}=\mathrm{min}\left(\frac{\rho}{\rho_{\mathrm{SF}}},0.2\right).
\end{eqnarray} 

The merging of BH pairs occurs when their separations fall below the smoothing length of the BHs; this is the minimum radius of a sphere that encloses a specified number of neighboring gas cells weighted over a smoothing kernel. Due to the limited resolution, the small-scale dynamics of BHs cannot be determined self-consistently; this is particularly true when the BH mass is smaller than the mass of the DM particles. To avoid spurious forces, the BHs are therefore re-positioned to the location of the closest potential minimum.

\subsection{Modelling of black hole seeds}
\label{Modelling of black hole seeds}

Our seed models are based on the gas properties of halos and are designed to emulate conditions for DCBH seed formation. We first apply a set of seeding criteria to restrict the seeding to halos with gas that is metal poor and has a density above the minimum threshold for star formation; we hereafter refer to this as ``dense, metal poor gas". In particular, seeds of mass $M_{\mathrm{seed}}$ are allowed to form only in halos which satisfy:

\begin{itemize}
    \item a minimum threshold for dense, metal poor gas mass, denoted by $\tilde{M}_{\rm sf,mp}$. As in \cite{2021arXiv210508055B}, the tilde indicates that the mass threshold is a dimensionless quantity normalized to the seed mass: $\tilde{M}_{\rm sf,mp} \equiv {M}_{\rm sf,mp}/M_{\rm seed}$. `Metal poor' gas cells refer to those with metallicities less than $10^{-4}~Z_{\odot}$. Note however that our results are not significantly sensitive to the choice of this threshold from $10^{-5}-10^{-2}~Z_{\odot}$.   
    \item a minimum threshold for the total mass, denoted by $\tilde{M_h}$~(this is also a dimensionless quantity normalized to the seed mass: $\tilde{M}_{\rm h} \equiv {M}_{\rm h}/M_{\rm seed}$).
\end{itemize}

A range of models with the above seeding criteria~(in the parameter space of $\tilde{M}_{\mathrm{sf,mp}}$, $\tilde{M}_{h}$ and $M_{\mathrm{seed}}$) has been explored in \cite{2021arXiv210508055B}, where we found that both $\tilde{M}_{\mathrm{sf,mp}}$ and $\tilde{M}_{h}$ leave strong and distinct imprints on the merger rates, and therefore also the BH masses. More specifically, a factor of 10 increase in $\tilde{M}_{\mathrm{h}}$ causes $\sim100$ times suppression of merger rates at $z>15$; $\tilde{M}_{\mathrm{sf,mp}}$ has greater impact at lower redshifts~($z\sim7-15$), where it can suppress the merger rates by factors of $\sim8$ when increased from 5 to 150.

In this work, we fix $\tilde{M}_{h}=3000,~\tilde{M}_{\mathrm{sf,mp}}=5$ and explore seed masses of $M_{\mathrm{seed}}=1.25\times10^{4},1\times10^{5},~8\times10^5~M_{\odot}/h$; this is motivated by a number of considerations. 
First, the seed masses and the corresponding halo masses for their formation~($\sim10^7-10^9~M_{\odot}/h$ halos) are consistent with theoretical predictions for where DCBHs are expected to form~\citep{2003ApJ...596...34B,2004MNRAS.354..292K}. Second, the choice of $\tilde{M}_{h}=3000,~\tilde{M}_{\mathrm{sf,mp}}=5$ provides reasonably well converged results with respect to increasing resolution~\citep{2021arXiv210508055B}. Third, this model also produces a sufficient number of BHs within our zoom volume~(to be descrbed in Section \ref{Simulation suite}), so that we can put in additional criteria to further restrict the seeding and investigate their impact. 
Hereafter, we shall refer to the above criterion as the \textit{baseline seeding criteria}. 

Having applied the baseline seeding criteria, we then explore the impact of further restricting the seeding based on the gas angular momentum and LW flux as described in the following subsections.

\subsubsection{Gas spin criterion}

Here, we restrict the seeding to halos with low gas angular momentum. For each halo, we compute the net angular momentum of all gas cells with respect to their center of mass, which we hereafter refer to as ``gas spin"~($\vec{\mathbf{J}}_{\mathrm{spin}}$), as
\begin{equation}
    \vec{\mathbf{J}}_{\mathrm{spin}}= \sum^{\mathrm{gas~cells}}_{i} \left[ \vec{\mathbf{r}}_i\times\vec{\mathbf{p}}_i - \vec{\mathbf{r}}_{\mathrm{com}}\times\vec{\mathbf{p}}_{\mathrm{com}} \right] 
\end{equation}
where the summation is over all gas cells around the halo potential minimum up to the halo virial radius $R_{\mathrm{vir}}$; $\vec{\mathbf{r}}_i$ and $\vec{\mathbf{p}}_i$ are the position and momentum of the $i^{th}$ gas cell. $\vec{\mathbf{r}}_{\mathrm{com}}$ and $\vec{\mathbf{p}}_{\mathrm{com}}$ are the position and momentum of the center of mass of gas cells within the virial radius. We define the dimensionless gas spin parameter of the halo as
\begin{equation}
    \lambda =\frac{|\vec{\mathbf{J}}_{\mathrm{spin}}|}{\sqrt{2}M_{\mathrm{gas}} R_{\mathrm{vir}} V_{\mathrm{vir}}}
    \label{lambda_eqn}
\end{equation}
where $M_{\mathrm{gas}}$ is total gas mass within the halo virial radius, and $V_{\mathrm{vir}}=\sqrt{\frac{GM_{\mathrm{vir}}}{R_{\mathrm{vir}}}}$ is the circular velocity. 

Our gas spin criterion is motivated by the results of \cite{2006MNRAS.371.1813L} on the stability analysis of pre-galactic gas discs in high redshift halos~(as also adopted by \citealt{2012MNRAS.422.2051N} and \citealt{2020MNRAS.491.4973D}). They derive a  maximum gas spin $\lambda_{\mathrm{max}}$, above which the gas disc is gravitationally stable. At lower gas spins, the disc becomes prone to gravitational collapse, potentially resulting in a massive DCBH seed. This maximum gas spin is given by
\begin{equation}
  \lambda_{\mathrm{max}}=\frac{m_d^2 Q_c}{8 j_d} (T_{\mathrm{vir}}/T_{gas})^{1/2} 
\end{equation}
where $m_d$ and $j_d$ are the fractions of the mass and angular momentum respectively, of the halo that forms the disc. $Q_c$ is the Toomre instablity parameter. $T_{\mathrm{vir}}$ is the virial temperature of the halo and $T_{\mathrm{gas}}$ is the mean gas temperature. When $\lambda<\lambda_{\mathrm{max}}$, the fraction of the disk mass that falls towards the center~(providing fuel for BH seed formation) is $\sqrt{1-\lambda/\lambda_{\mathrm{max}}}$. 

We now focus on the implications of the foregoing physical arguments on our seed models. For a given seed mass $M_{\mathrm{seed}}$ to form, halos must have: 1) a gravitationally unstable disk and 2) a sufficient amount of gas mass~($>M_{\mathrm{seed}}$) collapsing to the center as a consequence. This corresponds to the following seeding criteria:
\begin{equation}
    \lambda<\lambda_{\mathrm{max}}
\label{gas_spin_criterion_eqn}
\end{equation}
and
\begin{equation}
    M_h > \frac{M_{\mathrm{seed}}}{m_d
    \sqrt{1-\lambda/\lambda_{\mathrm{max}}}}
\label{gas_spin_criterion_halo_eqn}
\end{equation}
where $M_h$ is the halo mass. Eq.~(\ref{gas_spin_criterion_halo_eqn}) is essentially derived by inverting Eq.~(1) from \cite{2012MNRAS.422.2051N}; as it turns out, it corresponds to halo mass thresholds which typically lie between $\sim100-500~M_{\mathrm{seed}}$, which is much smaller than our baseline criteria for halo mass~($\tilde{M}_h=3000$). Therefore, it is only Eq.~(\ref{gas_spin_criterion_eqn}) that impacts our seeding. 

To determine $\lambda_{\mathrm{max}}$ we compute $T_{\mathrm{vir}}$ and $T_{\mathrm{gas}}$ for every halo on the fly during the simulation. The parameters $m_d$, $j_d$ and $Q_c$ can depend on the structure of disks, which may not be well resolved for all of our simulations, particularly in low mass halos at early epochs. Computing these quantities on the fly would also be computationally demanding. For these reasons we instead simplify our model by assuming $m_d=j_d=0.05$ and $Q_c=2$, as also done in \citealt{2012MNRAS.422.2051N} and \cite{2020MNRAS.491.4973D}. We test the choices for $m_d$ and $j_d$ using one of our highest resolution simulations~(gas mass resolutions $\sim10^3~M_{\odot}/h$), where we compute them in post-processing for our seed forming halos using a kinematic decomposition of gas cells~(as done in \citealt{2018MNRAS.478.5063H}); the values tend to lie between 0.01 to 0.1, broadly consistent with our assumed value of 0.05. Hereafter, we shall refer to Eq.~(\ref{gas_spin_criterion_eqn}) as the \textit{gas spin criterion}.

\subsubsection{Lyman Werner (LW) flux criterion}
\label{Lyman Werner (LW) flux criterion}

We also examine the impact of restricting the seeding to halos exposed to a LW flux above a critical threshold. We first describe our methodology to compute the LW flux over the entire simulation box. Note that our simulations do not include direct radiative transfer. Therefore, we adopt an empirical prescription to compute the LW flux on the fly. In principle, the LW flux at a given location consists of a background component~(originating from distant stars not necessarily within the simulation volume) and a spatially varying component~(originating from nearby stars). The background component depends on the global star formation rate density, and is estimated to be $0.01~J_{21}$ at $z\sim25$ to $1~J_{21}$ at $z\sim7-10$~\citep{2013MNRAS.428.1857J}. The spatially varying component can however be much higher than the background component, particularly at the sites of potential seed formation. Moreover, the flux thresholds we plan to consider are also $\sim10-300$ times higher. Therefore, we neglect the background component and include only the spatially varying component in our calculation~(as also done in \citealt{2016MNRAS.463..529H}). 

The spatially varying components for both Pop III and Pop II LW fluxes~(adopted from \citealt{2014MNRAS.442.2036D} and \citealt{2021MNRAS.503.5046L}) are given by
\begin{equation}
J_{\mathrm{LW}}=\sum_i \frac{\left< h \nu \right>}{\Delta \nu} \frac{f_{\mathrm{esc}} Q_{LW}}{16 \pi^2 r_i^2} m_{*,i}
\label{LW_POPIII_eqn}
\end{equation}
where $m_{*,i}$ is the mass of each resolution element comprised of young stars, $r_{i}$ is the corresponding distance. $\nu=2.99\times10^{15}~\mathrm{Hz}$ and $\Delta \nu= 7.79\times10^{14}~\mathrm{Hz}$ are mean frequency and band width respectively of the LW band~($11.2-13.6~\mathrm{eV}$). $f_{\mathrm{esc}}$ is the escape fraction of LW photons, which is assumed to be 1 as done in both \citealt{2014MNRAS.442.2036D} and \citealt{2021MNRAS.503.5046L}. This assumption may not necessarily be true, in which case our calculated LW fluxes, and therefore the number of seeds formed, 
would only correspond to upper limits. $Q_{LW}$ is the photon production rate~(adopted from \citealt{2003A&A...397..527S}) and is given by
\begin{equation}
    Q_{LW}=Q_0\left(1+\frac{t_{LW}}{4~\mathrm{Myr}}\right)^{3/2}\exp{\left(-\frac{t_{LW}}{300~\mathrm{Myr}}\right)}
    \label{LW_POPIII_eqn_II}
\end{equation}
where $t_{LW}$ is the time elapsed after a star-burst and $Q_0=10^{47}~\mathrm{s}^{-1}M_{\odot}^{-1}$. We include contributions only from star formation within the previous 5 Myr~(consistent with \citealt{2012MNRAS.425.2854A,2014MNRAS.443..648A,2020MNRAS.491.4973D}), since most LW photons are emitted within this time interval due to the shorter lifetimes of the most massive Pop II and Pop III stars. Therefore, $m_{*,i}$ in Eq.~(\ref{LW_POPIII_eqn}) is assumed to be the total stellar mass formed in the last 5 Myr within the $i^{th}$ star forming gas cell, given by
\begin{equation}
    m_{*,i}=\mathrm{SFR}_i \times 5~\mathrm{Myr}
\end{equation} 
where $\mathrm{SFR}_i$ is the instantaneous star formation rate of the gas cell at a given time-step. In other words, each star forming gas cell is assumed to contain a single stellar population~(SSP) characterized by its age and metallicity. We only include gas cells representing Pop III and Pop II SSPs, which are classified by metallicities of $Z<0.001~Z_{\odot}$ and $0.001<Z<0.1~Z_{\odot}$, respectively. Star-forming gas with $Z>0.1~Z_{\odot}$~(Pop I stars) does not contribute to our Lyman Werner flux calculation since 1) they have redder spectra, and 2) they form a very small fraction~($\sim5\%$ at $z\sim11$ and $\lesssim1\%$ at $z\gtrsim16$) of the total stellar content in our zoom volume at such high redshifts. Lastly, note that our simulation time resolution is not high enough to resolve the detailed star formation history of each gas cell within 5 Myr interval; therefore, we simply assume that the entire stellar mass $m_{*,i}$ has an age of 5 Myr, and assign $t_{LW}=5~\mathrm{Myr}$ in Eq.~(\ref{LW_POPIII_eqn_II}).


\begin{table*}
     \centering
    \begin{tabular}{c|c|c|c|c|}
         $L_{\mathrm{max}}$ & $M_{dm}$~($M_{\odot}/h$) & $M_{gas}$~($M_{\odot}/h$) & $\epsilon~(kpc/h)$ & $M_{\mathrm{seed}}$~($M_{\odot}/h$) values explored\\
         \hline
         10 & $1\times10^6$ & $\sim10^5$ & 0.5 & $8\times10^{5}$, $1\times10^{5}$ \\
         11 & $1.3\times10^5$ & $\sim10^4$ & 0.25 & $8\times10^{5}$, $1\times10^{5}$, $1.25\times10^{4}$  \\
         12 & $1.6\times10^4$ & $\sim10^3$ & 0.125 & $8\times10^{5}$, $1\times10^{5}$, $1.25\times10^{4}$ \\
         \hline
    \end{tabular}
    \caption{Spatial and mass resolutions within the zoom region of our simulations for various values of $L_{\mathrm{max}}$~(see Section \ref{Simulation suite} for the definition). $M_{dm}$ is the mass of a dark matter particle, $M_{gas}$ is the typical mass of a gas cell~(note that gas cells can refine and de-refine depending on the local density) and $\epsilon$ is the gravitational smoothing length. The 4th column corresponds to the seed masses allowed at each $L_{\mathrm{max}}$, which is limited by the gas mass resolution. }
    \label{tab:my_label}
\end{table*}

For seeding BHs based on the calculated LW fluxes, we assume a threshold LW flux~(Pop II + Pop III contribution) required for seeding, denoted by $J_{\mathrm{crit}}$. The seeding criterion then requires that the dense, metal poor gas must also be illuminated by LW intensities $>J_{\mathrm{crit}}$~(hereafter referred to as ``LW illuminated"). This can then be expressed as
\begin{equation}
    M_{\mathrm{sf,mp,LW}} >  \tilde{M}_{\mathrm{sf,mp}}~M_{\mathrm{seed}}=5~M_{\mathrm{seed}}
    \label{LW_seed_criterion_eqn}
\end{equation}
where $M_{\mathrm{sf,mp,LW}}$ is the total mass of dense, metal poor gas cells within a given halo that are also exposed to LW fluxes greater than $J_{\mathrm{crit}}$. Lastly, we also make sure that star formation is switched off within these dense, metal poor, LW illuminated gas cells. We hereafter refer to this as the \textit{LW flux criterion}. Note that we continue to use the subscript `sf'~(which stands for `star forming') to be consistent with the notation in \cite{2021arXiv210508055B}; however, when the LW flux criterion is applied, `sf' corresponds gas with densities exceeding the star formation threshold but \textit{does not} actually form stars.


\subsection{Simulation suite}
\label{Simulation suite}
\begin{figure*}
\includegraphics[width=16 cm]{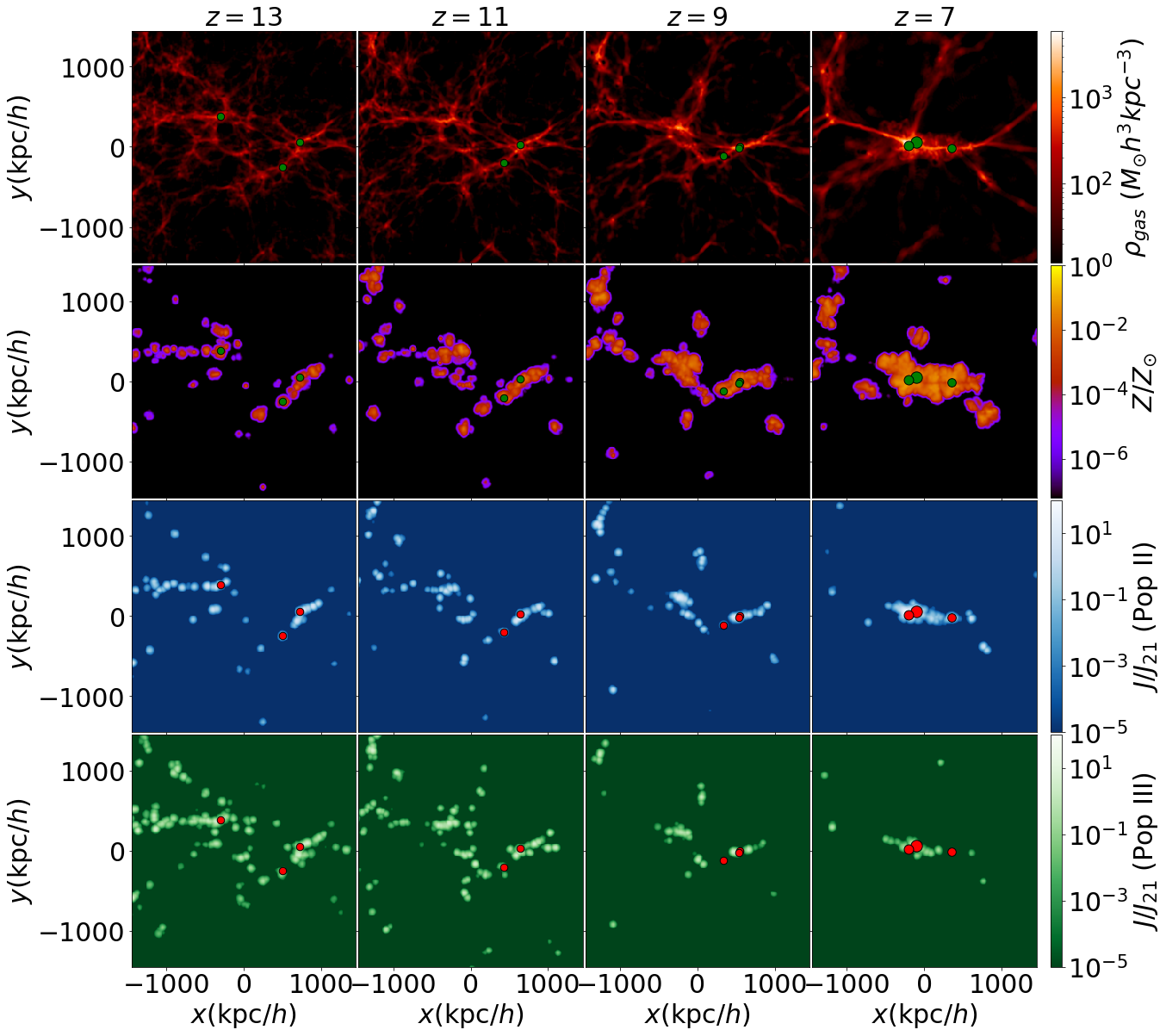}
\caption{2D projected color maps of the gas density~(1st row), gas metallicity~(2nd row) and LW fluxes from Pop II~(3rd row) and Pop III~(4th row) stars in our zoom region. Left to right panels show the redshift evolution from $z=13$ to $z=7$. The green and red circles represent BHs seeded at $1\times10^{5}~M_{\odot}/h$ in halos with $\tilde{M}_{h}=3000$, $\tilde{M}_{\mathrm{sf,mp}}=5$ and $J_{\mathrm{crit}}=50~J_{21}$. As time evolves, the star forming and metal enriched regions appear throughout our zoom volume. These regions are sources of LW photons. Seed black holes are formed in metal poor regions where gas densities exceed the star formation threshold~(hereafter referred as ``dense gas, metal poor gas"), but the star formation itself is suppressed by the Lyman Werner flux.}
\label{density_2dplot_fig}
\end{figure*}
Our simulation suite consists of a series of zoom simulations of a universe with an underlying cosmology adopted from \cite{2016A&A...594A..13P}, ~($\Omega_{\Lambda}=0.6911, \Omega_m=0.3089, \Omega_b=0.0486, H_0=67.74~\mathrm{km}~\mathrm{sec}^{-1}\mathrm{Mpc}^{-1},\sigma_8=0.8159,n_s=0.9667$). The initial conditions~(ICs) are generated using \texttt{MUSIC}~\citep{2011MNRAS.415.2101H} within a parent box with $(25~\mathrm{Mpc}/h)^3$ comoving volume. 

Our density field realization and the zoom-in region of interest is the same as that of \cite{2021arXiv210508055B}. Here we briefly summarize the main features and refer the interested reader to \cite{2021arXiv210508055B} for more details. We first ran a uniform volume simulation with $128^3$ particles and selected a target halo of mass $3.5\times10^{11}~M_{\odot}/h$~(corresponding to a peak height $\nu=3.3$)  at $z=5$ to resimulate at higher resolutions. DM particles comprising that halo were traced to $z=127$, wherein a cuboidal region enclosing these particles is selected for the zoom runs; this region was referred to as \texttt{ZOOM_REGION_z5} in \cite{2021arXiv210508055B}. However, note that for this work, we increased the dimensions of the initial~($z=127$) cuboidal zoom region by $50\%$ to allow for higher number of seeds to form in regions free of contamination from low resolution DM and gas particles. Therefore, our baseline model in this work produces $\sim2-3$ times higher number of seeds compared to that of \cite{2021arXiv210508055B}. 

For our zoom-in ICs, the resolution of the zoom region is characterized by the parameter $L_{\mathrm{max}}$, corresponding to a uniform box with $2^{L_{\rm max}}$ DM particles per side. Table \ref{tab:my_label} summarizes the mass and spatial resolutions in our zoom region for different values of $L_{\mathrm{max}}$. The background grid is always kept at $L_{\mathrm{min}}=7$, while for the zoom region we explored $L_{\mathrm{max}}=10,11,12$. In \cite{2021arXiv210508055B}, we found that for our baseline seeding model, the results were reasonably well converged by $L_{\mathrm{max}}=11$. Additionally, $L_{\mathrm{max}}=11$ also runs in reasonable enough time to be able to explore wide range of seeding parameters; therefore, we primarily use $L_{\mathrm{max}}=11$ for this study. However, we do find that the additional seeding criteria~
(particularly the LW flux criterion) can impact resolution convergence; we discuss this in more detail in Section \ref{Seeding at higher resolution zooms} and Appendix \ref{appendix_resolution_convergence}. 

\subsubsection{Black hole seed models explored}

Here we summarize the gas based seed models explored in this work. The key parameters of interest in our modelling are $\tilde{M}_{h},\tilde{M}_{\mathrm{sf,mp}},M_{\mathrm{seed}}~\&~J_{\mathrm{crit}}$. In addition, we have the gas spin criterion which can be switched on or off. As discussed in Section \ref{Modelling of black hole seeds}, $\tilde{M}_{h}$ and $\tilde{M}_{\mathrm{sf,mp}}$~(which were explored in \citealt{2021arXiv210508055B}) are kept fixed at $3000$ and $5$, respectively.

When exploring models with the LW flux criterion, we consider $J_{\mathrm{crit}}$ values $10,50~\&~100~J_{21}$. While these are substantially below current theoretical predictions for $J_{\mathrm{crit}}$ from hydrodyanamic simulations and one-zone chemistry models~($\gtrsim1000~J_{21}$), 
it is clear from our results that such high values of $J_{\mathrm{crit}}$ would not produce any seeds in our simulation volume. 
Therefore, we 
systematically explore several lower values of $J_{\mathrm{crit}}$ and use the results to understand potential implications for DCBH formation. In future work, we plan to explore higher $J_{\mathrm{crit}}$ values in larger, more overdense halos. The $M_{\mathrm{seed}}$ values explored in this work are $1.25\times10^4,1\times10^5~\&~8\times10^5~M_{\odot}/h$, which broadly span the masses up to which DCBHs 
are expected to form via runaway infall of gas in halos with virial temperatures $T_{\mathrm{vir}}\gtrsim10^4~K$~\citep{2006MNRAS.370..289B}.

We use the following nomenclature to refer to our models for the remainder of this work. If a particular model only applies the baseline seeding criteria, we label it is `\texttt{BASELINE}'. When the gas spin criteria is included, we label it as `\texttt{LOWSPIN}'. When the LW flux criterion is included, then we include `\texttt{LW*}' where `*' is replaced by the $J_{\mathrm{crit}}$ flux value. For example, $J_{\mathrm{crit}}=50~J_{21}$ would correspond to `LW50'. 


\section{Results}
\label{Results}

\subsection{Build up of seed formation sites}
\begin{figure}
\includegraphics[width=8 cm]{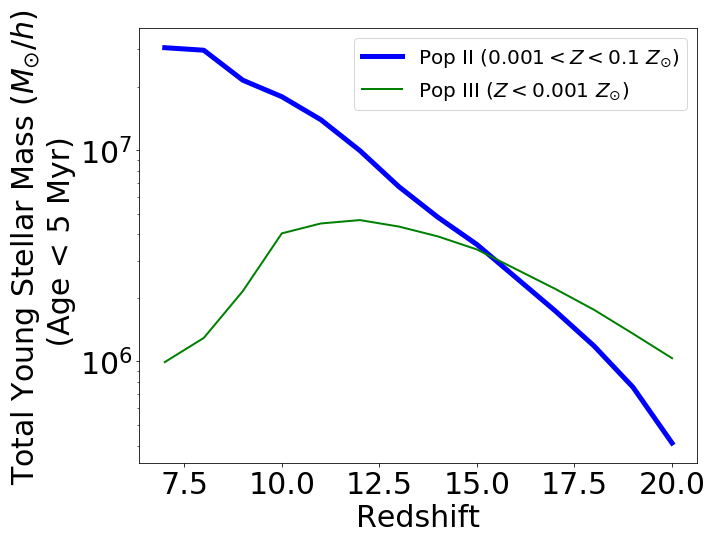}
\caption{Total mass of young stars~($\mathrm{age} < 5~\mathrm{Myr}$) in the zoom region for Pop III~($Z<0.001~Z_{\odot}$; green curve) and Pop II~($0.001<Z<0.1~Z_{\odot}$; blue curve) components. We find that between these components, Pop III stars dominate at $z\gtrsim15$ and Pop II stars dominate at $z\sim7-15$.}
\label{Young_stellar_mass}
\end{figure}

\begin{figure*}
\includegraphics[width=16 cm]{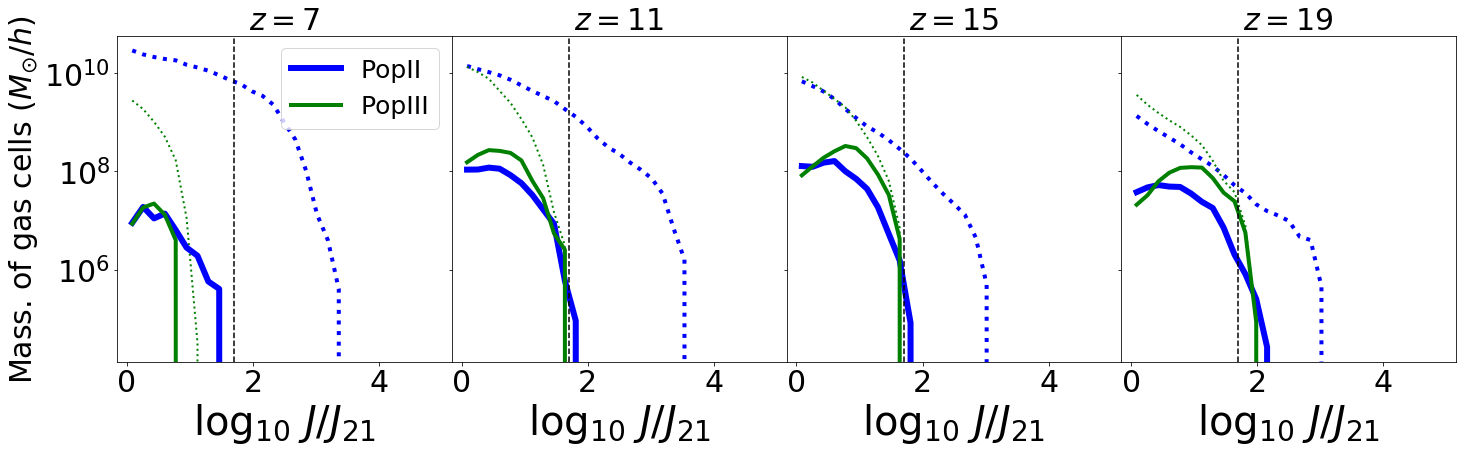}
\caption{Total mass of gas cells illuminated by LW photons originating from Pop II~(blue) and Pop III~(green) stars within bins of various flux values shown in the x-axis. Dotted lines correspond to all gas cells and solid lines correspond to dense, metal poor gas cells. Black vertical lines correspond to flux thresholds of $J_{\mathrm{crit}}=50~J_{21}$. We find that across all gas cells, Pop II stars are the dominant contributors to the LW radiation, with fluxes reaching up to a few $\times 10^3~J_{21}$. However, when we specifically look at only the dense, metal poor gas cells, the contribution from Pop II stars is substantially smaller and becomes comparable to that of Pop III stars at $z\gtrsim11$. Within dense, metal poor regions, flux values reach only up to $\sim100~J_{21}$.} 
\label{LW_intensity_popII_III}
\end{figure*}

Figure \ref{density_2dplot_fig} visualizes the evolution of the key properties of the gas distribution of the zoom region that drive the formation of seeds, which proceeds as follows. As time progresses~(Figure \ref{density_2dplot_fig} shows $z=13$ to $z=7$)  gravitational collapse and gas cooling leads to regions with densities high enough to trigger star formation~(Figure \ref{density_2dplot_fig}: 1st row). Subsequent stellar evolution processes lead to a significant amount of metal enrichment~(Figure \ref{density_2dplot_fig}: 2nd row). These earliest stages of star formation and metal enrichment regions are primarily comprised of young Pop III and Pop II stellar populations, which bombard nearby gas with LW radiation.

The LW fluxes from Pop II and Pop III stars are shown in the 3rd and 4th rows of Figure \ref{density_2dplot_fig}. The BH seeds start forming in regions illuminated by LW flux. However, these regions soon become metal enriched due to their close proximity to star forming gas; this stops the formation of new seeds. The dispersion of metals as well as LW photons are two competing processes that are simultaneously driven by star formation; as a result, the window for seed formation is relatively narrow. 

It is instructive to compare the impact of young Pop II vs. Pop III stars~($\mathrm{age} < 5~\mathrm{Myr}$) on seed formation in our models. Figure \ref{Young_stellar_mass} shows the total amount of young stellar content in the form of Pop II and Pop III stars, as a function of redshift. At $z\sim16-20$, the young stellar content is dominated by Pop III stars; this is because at this relatively early stage of star formation, a majority of the star forming regions have not yet been enriched by metals. However, by $z\sim7-15$, the gas is sufficiently enriched and Pop II stars start to dominate the young stellar population.

Figure \ref{LW_intensity_popII_III} compares the contributions from Pop III vs. Pop II stars to the LW fluxes on the surrounding gas. We first look at LW fluxes for all gas cells~(dotted lines in Figure \ref{LW_intensity_popII_III}); we find that at $z=7,11,15$, the LW flux is predominantly contributed by Pop II stars for all flux values between $\sim1$ to $\sim10000$; this is expected from the results of Figure \ref{Young_stellar_mass} where Pop II stars dominate the total stellar content at $z\sim7-15$.  Notably, even at $z\sim19$ where Pop III stars are more abundant overall, LW fluxes from Pop II stars still dominate at the highest values~($\gtrsim50~J_{21}$) relevant for BH formation. This is because the highest LW fluxes naturally occur in the densest regions, where metal enrichment~(and therefore, Pop II star formation) is expected to be more prevalent compared to other locations. 

Next, we look at the LW fluxes in gas cells that are simultaneously dense and metal poor~(solid lines in Figure \ref{LW_intensity_popII_III}). The first thing to note is that the LW flux contributed by Pop II stars is substantially smaller in dense, metal poor gas cells compared to all gas cells~(solid vs dotted blue lines in Figure \ref{LW_intensity_popII_III}). In contrast, the incident PopIII LW flux is similar for all gas and for the subset of dense, metal-poor gas cells, at least for LW fluxes $\gtrsim50~J_{21}$ 
(solid vs dotted green lines in Figure \ref{LW_intensity_popII_III}). These results are not unexpected since Pop II stars are likely to be somewhat further away from metal poor regions by construction~(recall that ``metal poor" implies $Z<10^{-4}~Z_{\odot}$, whereas Pop II stars have $10^{-3}<Z <10^{-1}~Z_{\odot}$). As a result, in dense, metal poor regions the difference between LW fluxes for Pop II vs Pop III stars is not drastically different, and both populations play an equally important role in seed formation. Lastly, the maximum LW fluxes in dense, metal poor regions are only up to a few times $\sim100~J
_{21}$; this already indicates that a seeding criterion of $J_{\mathrm{crit}}\gtrsim1000$ 
would not produce any seeds within \texttt{ZOOM_REGION_z5}. Therefore, the remainder of the paper will largely focus on significantly lower values of $J_{\mathrm{crit}}$ 
$ = 10-100~J_{21}$ and their impact on seed formation.  

\subsection{Characterizing halo properties relevant for seed formation}

\label{Characterizing halo properties relevant for DCBH formation}

\begin{figure*}
\includegraphics[width=16cm]{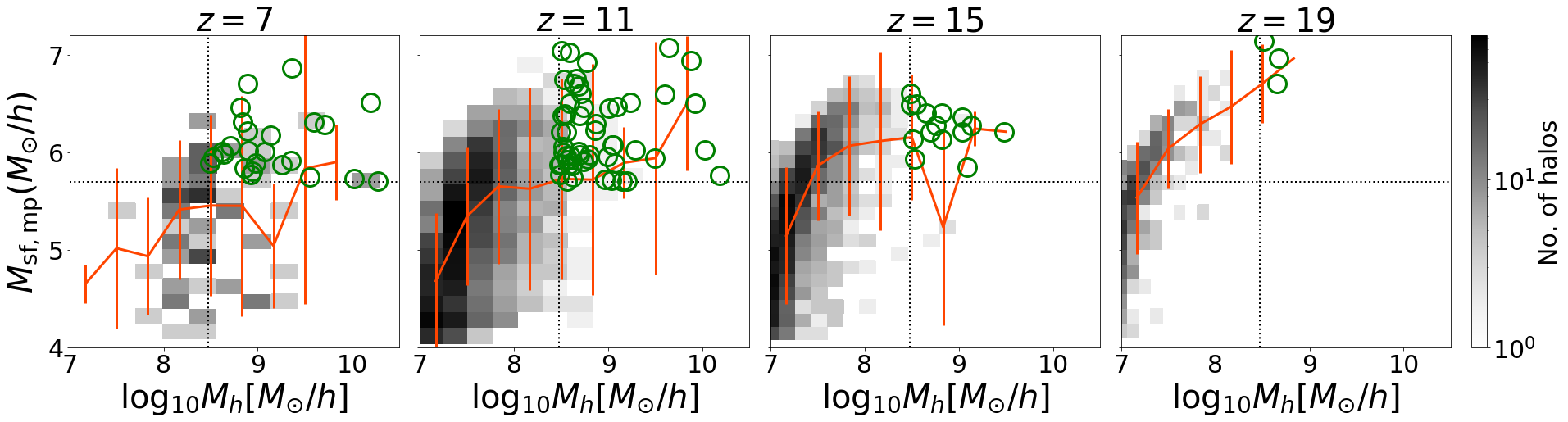} 
\includegraphics[width=16cm]{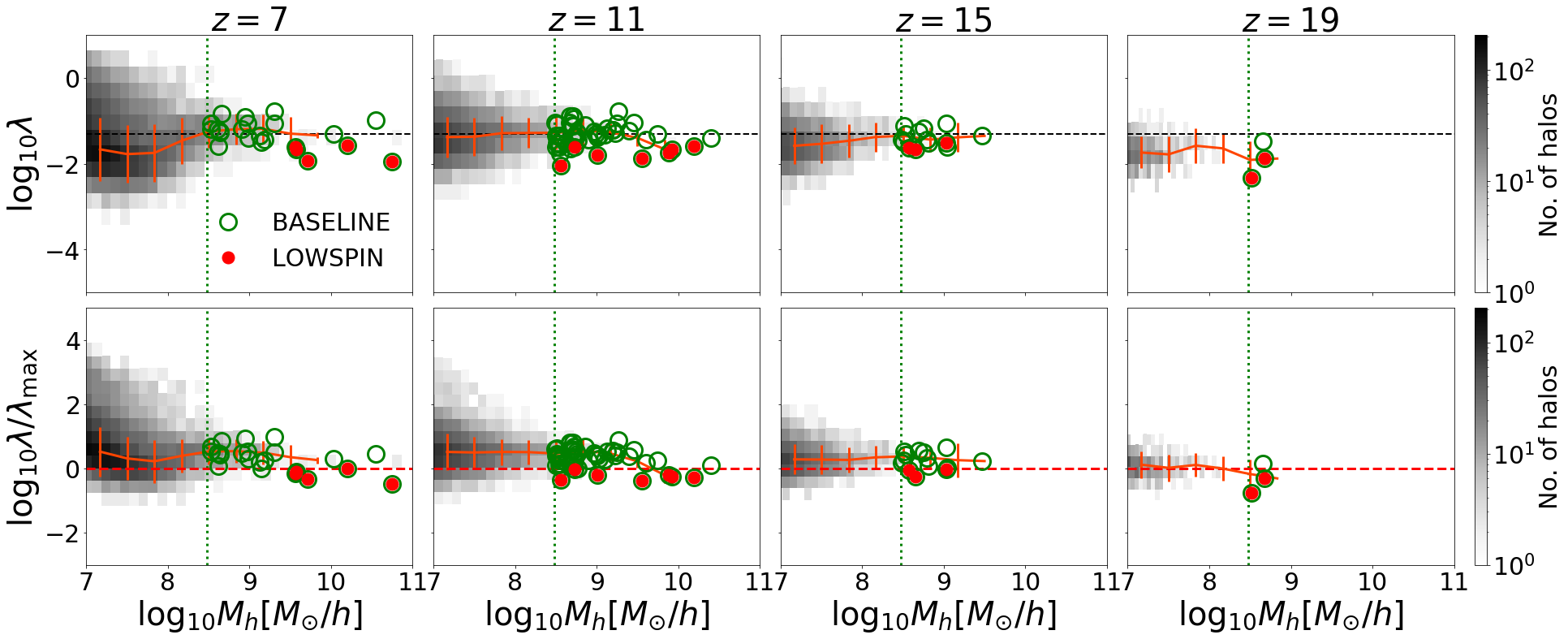} 
\includegraphics[width=16cm]{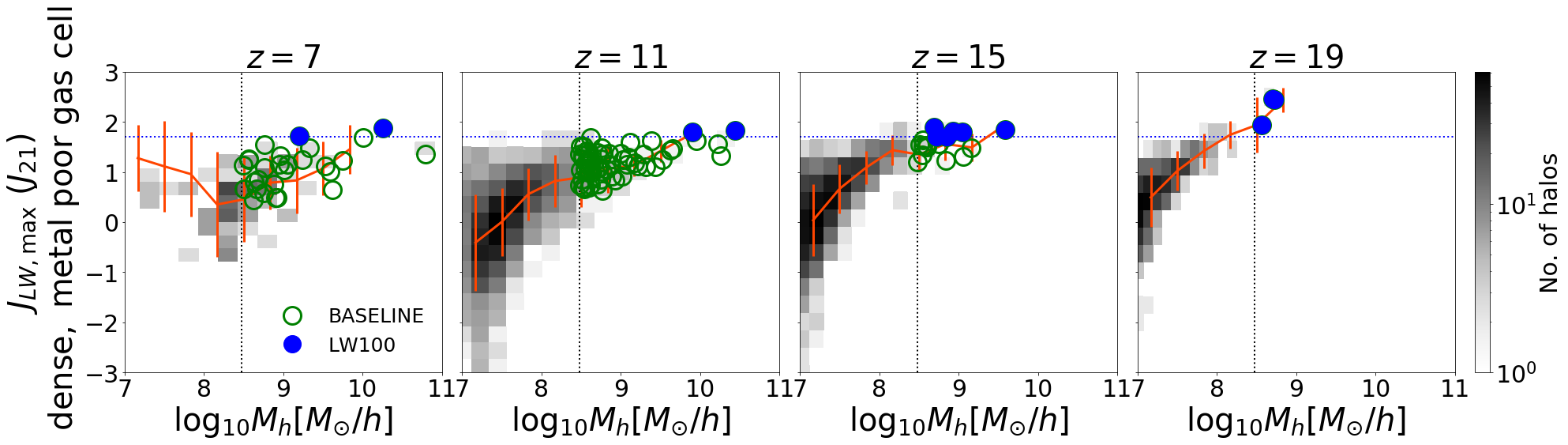} 
\caption{The relationships between the various halo properties that determine the formation of DCBHs at different redshifts from $z=7$ to $z=19$. We show only halos that have $<1\%$ contamination from low resolution dark matter particles. The 1st row shows halo mass vs. star forming, metal poor gas mass. The 2nd row shows the halo mass~($M_{h}$) vs. dimensionless gas spin~($\lambda$). In the 3rd row, the dimensionless spins from the 2nd row are normalized with respect to the maximum value~($\lambda_{\mathrm{max}}$) allowed for seeding to occur. The 4th row shows the halo mass vs the maximum LW flux amongst all dense, metal poor gas cells of the halo. The vertical lines are the minimum halo mass for seeding~($\tilde{M}_{h}=3000$). The horizontal line in the 1st row is the minimum dense, metal poor gas mass for seeding~($\tilde{M}_{sf,mp}=5$). In the 3rd row, the horizontal line is the maximum gas spin~($\lambda_{\mathrm{max}}$) that is allowed for seeding. In the 4th row, the horizontal line corresponds to $J_{LW}=50~J_{21}$. Green open circles are halos that satisfy the baseline seeding criteria~($\tilde{M}_{h}=3000$, $\tilde{M}_{\mathrm{sf,mp}}=5$). The red filled circles are halos that satisfy the baseline criteria as well as the gas spin criterion~($\lambda < \lambda_{\mathrm{max}}$). The blue filled circles are halos that satisfy the baseline criteria as well as the LW flux criterion~($M_{\mathrm{sf,mp,LW}}>5 ~M_{\mathrm{seed}}$; $J_{\mathrm{crit}}=50~J_{21}$). We find that only a small fraction of halos which satisfy the baseline seeding criteria also satisfy the gas spin and LW flux criterion.}
\label{Halo_properties_fig}
\end{figure*}

\begin{figure*}
\includegraphics[width=17 cm]{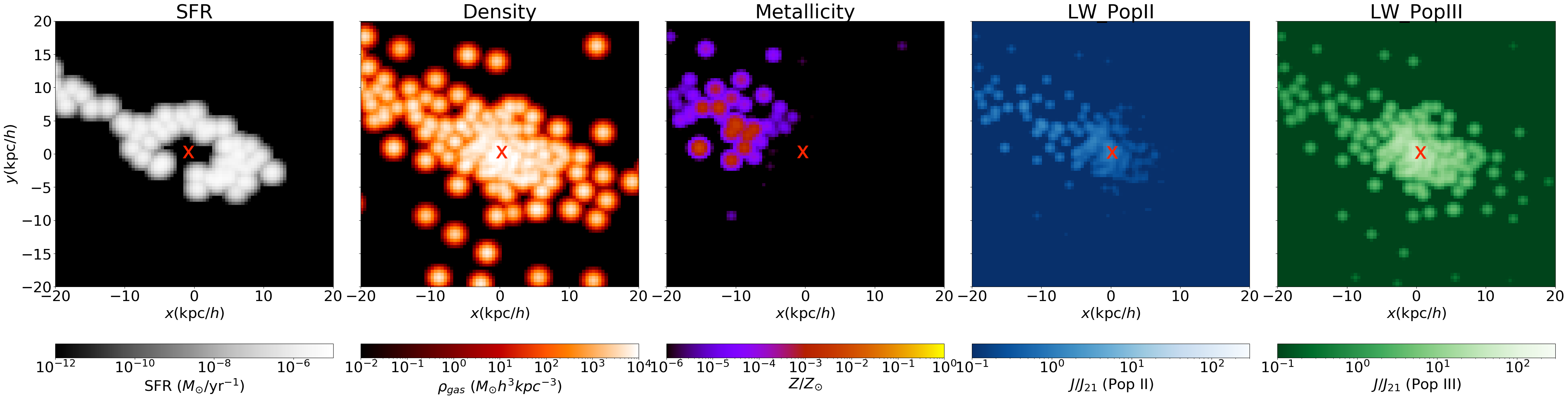}
\includegraphics[width=17 cm]{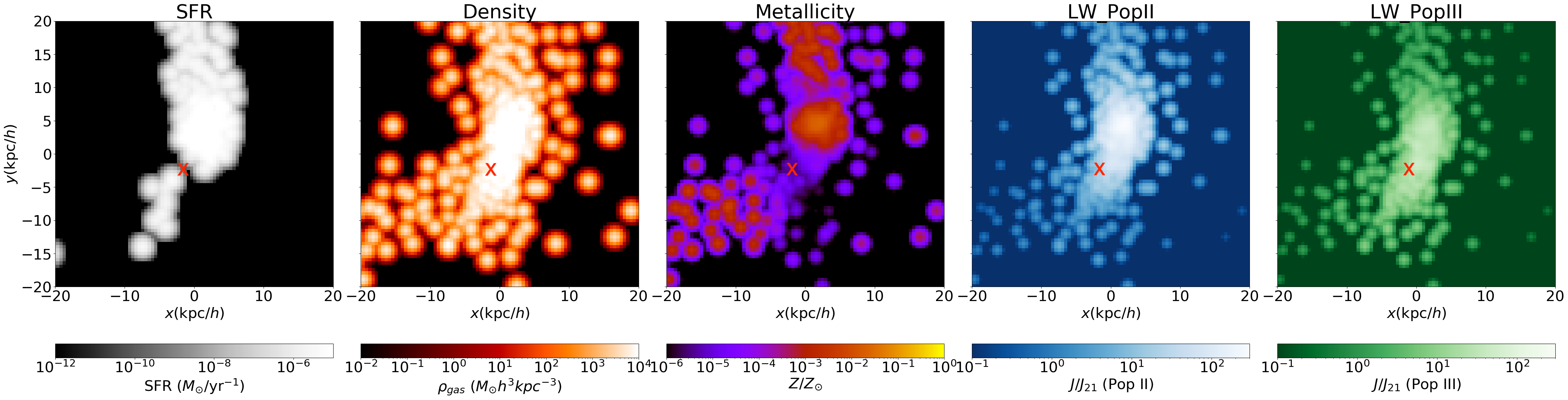}
\caption{Top and bottom panels are visualizations~(2D projected plots) of two different sites for BH seed formation that typically reside within halos of mass $\sim10^8-10^{10}~M_{\odot}/h$ i.e. a metal poor pocket~(see red cross) where the gas density exceeds the star formation threshold but the star formation is suppressed by LW radiation. The thickness of the slices along the line of sight is $5~\mathrm{kpc}/h$. At each pixel, the average value of the field is computed, followed by smoothening using a Gaussian filter of a fixed width at all locations. These are simulated at $L_{\mathrm{max}}=11$. From left to right, we show the star formation rate, gas density, gas metallicity, Lyman Werner fluxes from Pop III and Pop II stars. In this case, star formation in metal poor pockets is suppressed by LW fluxes greater than $50~J_{21}$.}
\label{Visulation_regions_fig_I}

\includegraphics[width=17 cm]{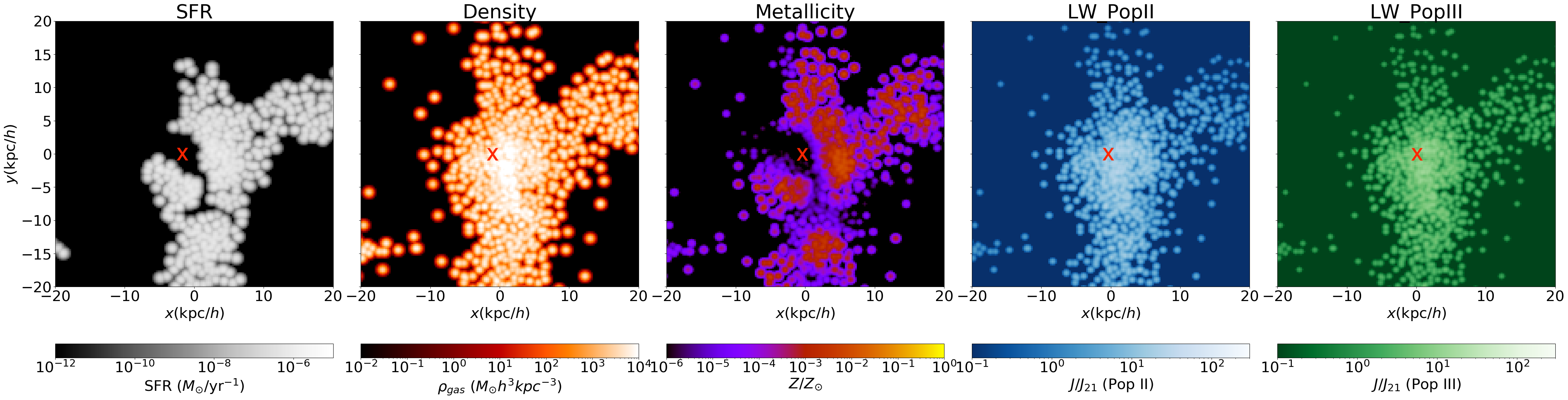}
\includegraphics[width=17 cm]{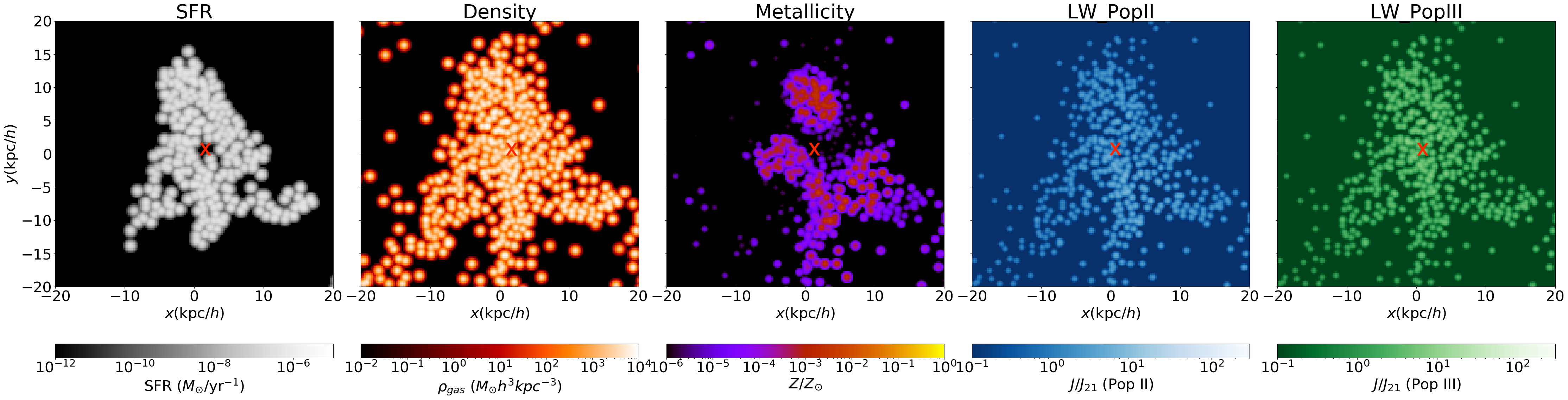}

\caption{Similar to the previous figure, the top and bottom panels here also show visualizations of two different sites of seed formation, but for $L_{\mathrm{max}}=12$ resolution. Dense, metal poor pockets also form at higher resolutions despite the higher rates of metal enrichment.} 
\label{Visulation_regions_fig_II}
\end{figure*}

\begin{figure*}
\includegraphics[width=5.5cm]{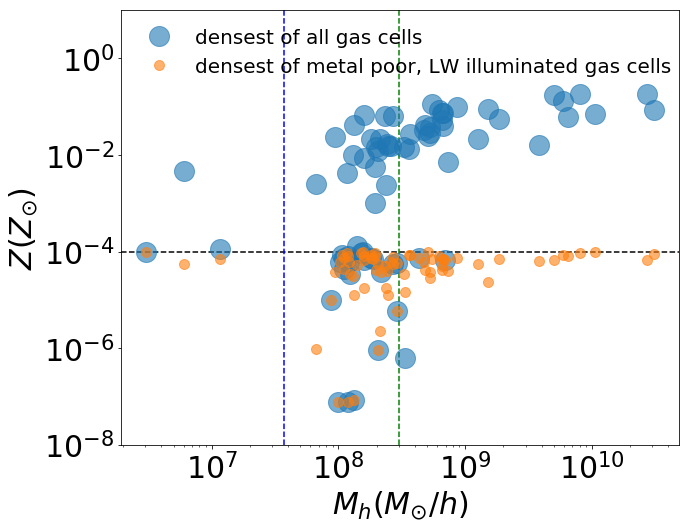} 
\includegraphics[width=5.5cm]{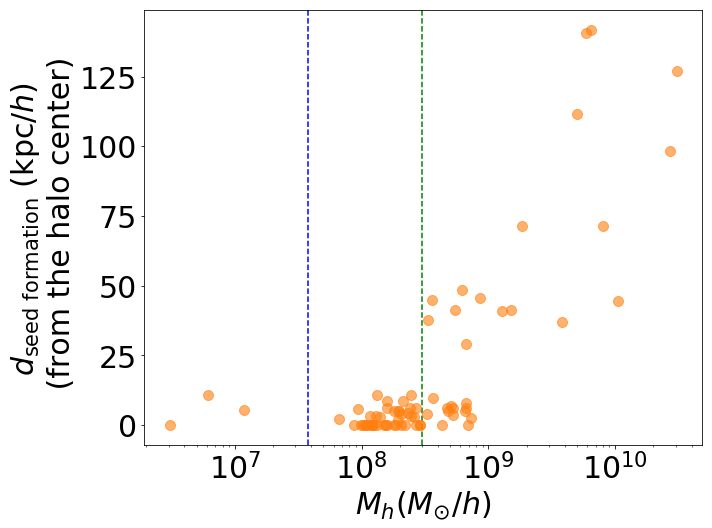} 
\includegraphics[width=5.5cm]{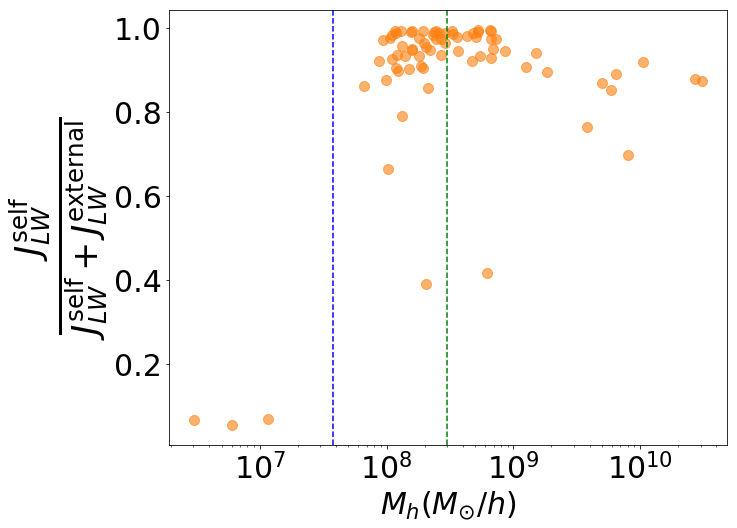} 
\caption{The scatter plots show key quantities for 
all halos that contain a non-zero amount of dense, metal poor, LW illuminated gas for all snapshots between from $z=7$ to $z=20$. \textbf{Left panel}: Gas-phase metallicity versus halo mass. Blue circles show the metallicity of the densest gas cell in a halo. Orange circles show the metallicity of the densest metal poor, LW illuminated gas cell in a halo. Blue and green dashed vertical lines correspond to halo mass thresholds for forming $1.25\times10^4~M_{\odot}/h$ and $1\times10^5~M_{\odot}/h$ seeds, respectively. The black horizontal dashed line is the metallicity ceiling for seed formation. Most seeds form at $\sim10^8-10^{10}~M_{\odot}/h$ halos wherein the densest metal poor gas cell is not at the halo center~(blue circles are significantly above orange circles); therefore, seed formation does not occur at the center of the halo. \textbf{Middle panel}:  Off-center distance of seed formation site versus halo mass. $d_{\mathrm{seed~formation}}$ is defined to be the distance between the site of seed formation~(densest metal poor, LW illuminated gas cell) from the halo center. For seed formation occurs increasingly further away from the halo center for increasingly massive halos. Seed formation can occur at distances up to $\sim130~\mathrm{kpc}/h$ from the halo center, but most seeds in our simulation form within 20 kpc/h of the halo center. \textbf{Right panel}: The ratio between the LW flux contributed by star forming gas present within the halo~($J_{LW}^{\mathrm{self}}$) vs the total LW flux that also includes star forming gas from outside the halo~($J^{\mathrm{self}}_{LW}+J^{\mathrm{external}}_{LW}$). Majority of the seed formation sites receive most~($\gtrsim90\%$) of the LW radiation from within the same halo.}
\label{seed_forming_pockets_fig}
\end{figure*}

In this section, we look at the $z\gtrsim7$ halo population in the zoom region and characterize it in terms of properties that are relevant for seed formation. In particular, we consider the total mass, dense~\&~metal poor gas mass, gas spin, and LW fluxes of halos in Figure \ref{Halo_properties_fig}. Note here that we only show halos where $\lesssim1\%$ of the total mass is contaminated by low resolution DM particles.

We first focus on how these different halo properties correlate with the halo mass for the overall population. The first row in Figure \ref{Halo_properties_fig} shows that the dense, metal poor gas mass positively correlates with the halo mass, particularly at $z\gtrsim 11$ (at $z=7$, there are too few metal poor halos in our volume to make definitive conclusions). This is not unexpected given that more massive halos also have higher gas content overall and typically have higher gas densities at their potential minima. That being said, metal enrichment will also be more prevalent in more massive halos~(due to the onset of star formation and evolution), thereby weakening the correlation; we can clearly see this happening for the most massive halos at $z\gtrsim11$. But despite the metal enrichment in these most massive star forming halos~($\sim10^8-10^{10}~M_{\odot}/h$), we still find that they have enough dense, metal poor gas mass to be potential sites of seed formation. As we shall see in Section \ref{Sites of seed formation in halos: Dense, metal poor, LW illuminated pockets}, the metal poor gas typically resides in pockets embedded within surrounding star forming and metal enriched~($\gtrsim10^{-4}~Z_{\odot}$) regions. 

The dimensionless gas spin~(shown in 2nd and 3rd rows of Figure \ref{Halo_properties_fig}) does not strongly correlate with halo mass. The gas spins are similar to those of the underlying dark matter spins, with mean values close to $\sim0.03-0.05$ at all halo masses and redshifts. These results are consistent with previous work using N-body simulations~\citep{2001ApJ...555..240B,2007MNRAS.378...55M,2007MNRAS.376..215B,2010MNRAS.404.1137B} as well as hydrodynamic simulations~\citep{2015MNRAS.449.2087D,2017MNRAS.466.1625Z,2020MNRAS.491.4973D}. In addition, the lack of halo mass vs. spin correlation is also a natural prediction from tidal torque theory~(see review by \citealt{2009IJMPD..18..173S}).

The last row of Figure \ref{Halo_properties_fig} shows that more massive halos are exposed to higher LW fluxes within their dense, metal poor gas. This is because more massive halos typically have higher amounts of star formation overall, and most of the LW radiation is coming from star forming regions within the same halo. As it turns out~(see section \ref{Sites of seed formation in halos: Dense, metal poor, LW illuminated pockets} for more detail), these halos contain pockets of dense, metal poor gas embedded within the star forming regions that provide the LW flux. Additionally, for halos at fixed mass, the LW flux typically decreases with time. This is simply due to Hubble expansion, which causes a star forming halo of a fixed mass to be less compact~(in physical coordinates) at lower redshifts, thereby leading to smaller distances between the star forming gas and the dense, metal poor pockets. Therefore, the formation of seeds in the presence of a LW flux criterion will be driven by two competing effects: 1) formation of more massive halos and proliferation of star forming regions with time, which will tend 
to increase the LW flux and form more seeds as redshift decreases, 2) Hubble expansion, which will tend 
to decrease the LW flux~(at fixed halo mass) and suppress the formation of seeds as redshift decreases. Recall also that these 
competing effects are in addition to two other pre-existing effects originating from the baseline criterion: namely, the 
formation of dense gas (which tries to increase the number of seeds at lower redshifts), and metal enrichment (which tries to decrease the number of seeds at lower redshifts). 

We now focus on the halo subsamples that satisfy different combinations of seeding criteria described in Section \ref{Modelling of black hole seeds}. The middle rows of Figure \ref{Halo_properties_fig} show that most halos satisfying the baseline seeding criteria do not satisfy the gas spin criterion~(filled red vs. open green circles). At $z=11$, for instance, only $\sim13\%$ of halos satisfying the baseline seeding criteria also satisfy the gas spin criterion. This implies that gas angular momentum should have a significant impact on seed formation. Next, we see in the bottom row of Figure \ref{Halo_properties_fig} that an even smaller fraction of halos satisfying the baseline criterion, also satisfy the LW flux criterion~(filled blue vs open green circles) with $J_{\mathrm{crit}}=50~J_{21}$. This suggests that the LW flux criterion may be even more stringent than the gas spin criterion. Note also that amongst the $z=$ 7, 11, 15, \& 19 snapshots shown, the $z=11$ snapshot has the highest number of halos satisfying either the baseline criterion or the gas spin criterion; but when the LW flux requirement is imposed, the peak epoch is at $z=15$.
This suggests that the LW flux criterion will push the peak of seed formation to higher redshifts compared to the baseline criterion. 
While Figure \ref{Halo_properties_fig}
shows the results for $M_{\mathrm{seed}}=1\times10^5~M_{\odot}/h$, the same inferences hold for all seed masses considered in this work.

\subsubsection{Sites of seed formation in halos: Dense, metal poor, LW illuminated pockets}
\label{Sites of seed formation in halos: Dense, metal poor, LW illuminated pockets}

Figure \ref{Visulation_regions_fig_I} shows visualizations of two different 
seed forming regions~(at our fiducial resolution $L_{\mathrm{max}}=11$) as projected 2D color plots of the star formation rate~(SFR), density, metallicity and LW fluxes from Pop II and Pop III stars. We can see that these regions have undergone substantial amounts of star formation and metal-enrichment, which is not surprising since they are significantly above the atomic cooling threshold. However, both halos contain small pockets~($\sim3-5~\mathrm{kpc}/h$; marked by red crosses) wherein the gas is still metal poor~($Z<10^{-4}~Z_{\odot}$). Additionally, the surrounding star forming regions provide LW flux to these pockets to completely suppress star formation, thereby creating an ideal site for seed formation. Note that the metal poor, LW illuminated  pockets are not located at the halo centers. Therefore, the seed formation in these halos will occur significantly away from the halo center. In our simulations, these seeds eventually end up at the halo center due to the BH repositioning scheme. However, recent simulations with more realistic treatment of BH dynamics have found that a substantial fraction of BHs may have difficulty in sinking to the halo centers, thereby leading to a population of off-center black holes even at low redshifts~\citep{2021,2021MNRAS.tmp.2452M,2021arXiv210209566B}. We shall investigate this in our simulations in future work. 

We also note that at higher resolutions~($L_{\mathrm{max}}=12$), the regions shown in Figure \ref{Visulation_regions_fig_I} no longer contain dense, metal pockets. This is because metal enrichment is not fully converged at $L_{\mathrm{max}}=11$; in particular, $L_{\mathrm{max}}=12$ has relatively earlier onset of metal enrichment~(see Figure 19 of \citealt{2021arXiv210508055B}). But nevertheless, dense metal poor pockets do also form at higher resolutions, as shown in Figure \ref{Visulation_regions_fig_II} for $L_{\mathrm{max}}=12$. Resolution convergence of the LW flux criterion is discussed further in Section \ref{Seeding at higher resolution zooms} and Appendix \ref{appendix_resolution_convergence}.

Next, we examine the formation of dense metal poor pockets in more detail for the full population of seed forming halos from $z\sim7-20$. In the left panel of Figure \ref{seed_forming_pockets_fig}, we show the metallicities at the halo centers~(more specifically the densest gas cell) of all seed forming halos~(for $J_{\mathrm{crit}}=50~J_{21}$) identified within snapshots from $z=7-20$. We find that for a significant majority of the seed forming halos between $\sim10^8-10^{10}~M_{\odot}/h$, the halo centers have metallicities of $\gtrsim10^{-2}~Z_{\odot}$. For these halos, the seed formation sites are not at the halo centers, and are located within dense, metal poor, LW illuminated pockets at distances that are mostly $\lesssim20~\mathrm{kpc}/h$, but can be up to $\sim130~\mathrm{kpc}/h$ from the halo center~(right panel of Figure \ref{seed_forming_pockets_fig}). These pockets have gas masses ranging from $\sim10^5-10^6~M_{\odot}/h$. Overall, this implies that in the presence of a LW flux criterion~($J_{\mathrm{crit}}=50~J_{21}$ or greater), the majority of seeds in our simulation are formed in the peripheral regions of $\sim10^8-10^{10}~M_{\odot}/h$ halos, instead of forming at the halo centers. These halos have a prior history of star formation and metal enrichment and for most of them, the dominant fraction~($\gtrsim~90\%$) of the LW radiation is contributed from their own star forming gas, and not from  neighboring halos. Notably, we also see that for seed formation sites in the most massive $\sim10^{10}~M_{\odot}/h$ halos, there is relatively higher contribution~($\sim20-30\%$) of LW radiation coming from neighboring halos. This is likely because in these halos, seed formation sites are farthest~($\gtrsim100$ kpc/h) from the central region of their host halos~(revisit middle panel of Figure \ref{seed_forming_pockets_fig}), thereby increasing their relative exposure to LW radiation from neighboring halos. 

The build up of seed formation sites in our simulations has some noteworthy distinctions compared to various models in the recent literature. For example~(as also mentioned in Section \ref{Introduction}), \cite{2017NatAs...1E..75R}, \cite{2021MNRAS.503.5046L} and  \cite{2014MNRAS.445.1056V} consider the formation of DCBHs via a pair of synchronised halos which cross the atomic cooling threshold within a few $\mathrm{Myr}$; the first halo to cross the threshold becomes star forming and provides LW radiation to another nearby halo. In this scenario, seeds would inevitably form in halos very close to the atomic cooling threshold with no prior history of star formation. Due to our model limitations~(lack of explicit $H_2$ cooling), we do not attempt to place seeds in halos very close to the atomic cooling threshold. Instead, we enforce a halo mass threshold for seeding~($\tilde{M}_h=3000$) which forces seeds to form in halos that have grown significantly since crossing the atomic cooling threshold~($1.25\times10^{4}~M_{\odot/h}$ seeds in $>3.7\times10^{7}~M_{\odot}/h$ halos and $1\times10^{5}~M_{\odot/h}$ seeds in $>3\times10^{8}~M_{\odot}/h$ halos).

Additionally, the LW flux criterion with $J_{\mathrm{crit}}/J_{21}=50~\&~100$ further enforces seed formation to largely occur at $\sim10^8-10^{10}~M_{\odot}/h$ halos~(revisit 4th row in Figure \ref{Halo_properties_fig}). Our simulations reveal that despite a prior history of star formation and metal enrichment in these halos, seeds can still form because the metals are not able to fully pollute the halo; this creates pockets of metal poor gas. If these pockets have dense gas that are also subjected to supercritical LW radiation from the surrounding star forming regions of the halo, they become sites of seed formation. This distinct DCBH formation scenario revealed by our simulations indicates that DCBHs may be slightly less rare and can be probed in somewhat smaller cosmological volumes than previously thought, while they still need to be significantly larger than our zoom volume~(particularly because of the high LW flux requirement i.e. $J_{LW}/ J_{21}\gtrsim1000$). Lastly, it is also noteworthy that despite the differences in seed formation scenario, there is one common implication between our models and those in the existing literature i.e. seeds are likely to end up in satellites of star-forming protogalaxies~\citep[see also][]{2014MNRAS.443..648A,2017ApJ...838..117N}.


\subsection{Impact of gas spin and LW flux on BH seeding}

\begin{figure*}
\includegraphics[width=16cm]{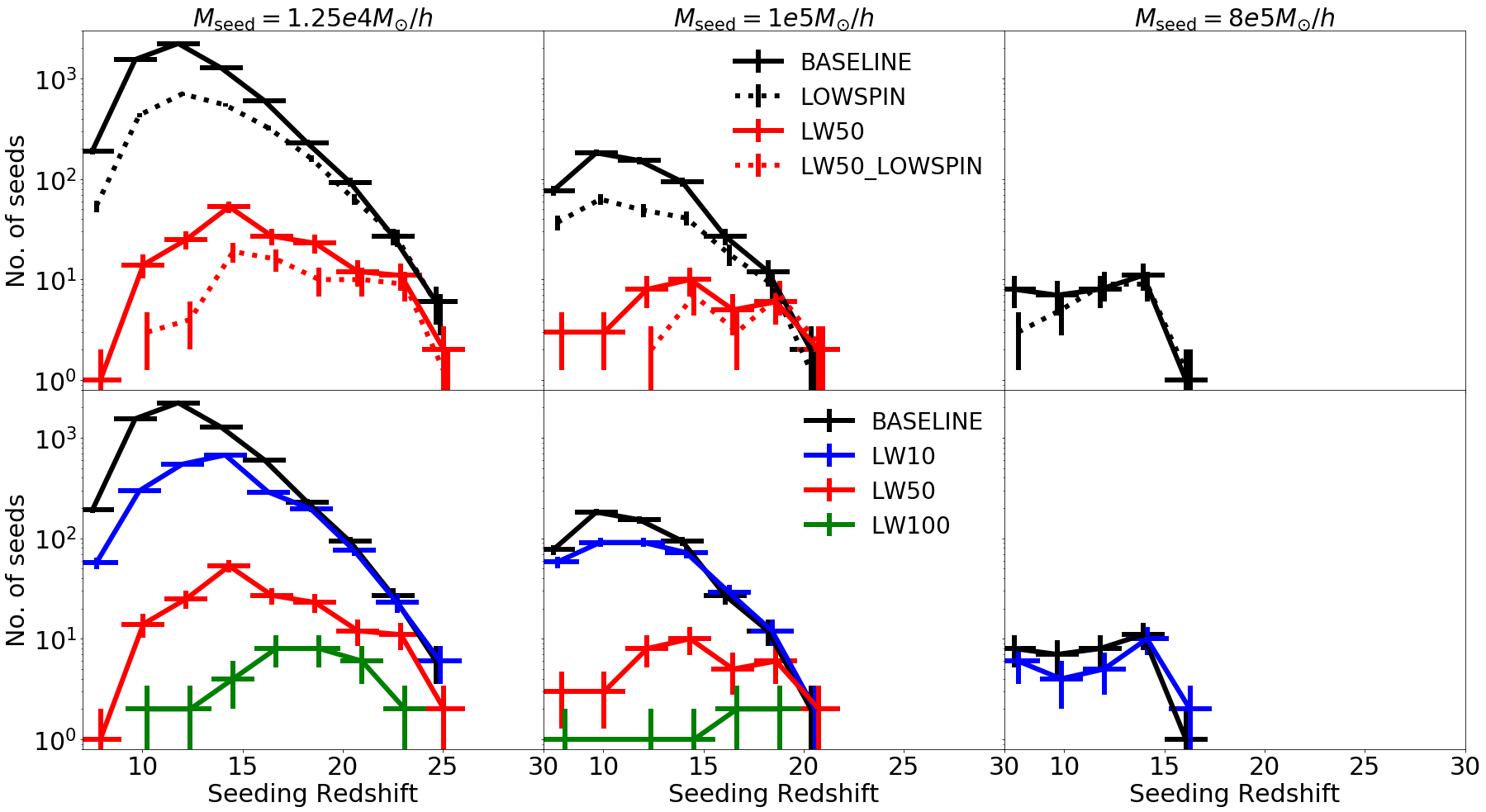} 
\caption{Distribution of seeding times for different seed models at fixed $M_{\mathrm{seed}}$. Dashed vs solid lines~(of the same color) in the upper panels correspond to models with vs. without the gas spin criterion, respectively. The suppression due to the gas spin criterion is by factors of $\sim6$ for all seed masses at $z\sim11-12$~(when most seeds form). Colored vs. black lines in the lower panels compare models with vs. without a LW flux criterion respectively. When a LW flux criterion with $J_{\mathrm{crit}}=50~\&~100~J_{21}$ is applied, $1.25\times10^4~M_{\odot}/h$ seeds are suppressed by factors of $\sim40~\&~300$ respectively; $1\times10^5~M_{\odot}/h$ seeds are suppressed by factors of $\sim20~\&~100$ respectively;  $8\times10^5~M_{\odot}/h$ seeds are completely suppressed.}
\label{impact_of_seeding_fig}
\end{figure*}

Here, we quantify the impact of gas spin and LW flux criteria on the frequency of BH seeding. Figure \ref{impact_of_seeding_fig} shows the number of seeds formed versus redshift, comparing the baseline model with the models in which a gas spin and/or LW flux criterion is added. Let us first focus on the gas spin criterion~(solid vs. dashed lines of same color in upper panels of Figure \ref{impact_of_seeding_fig}). At the highest redshifts~($z\gtrsim20$), adding the gas spin criterion does not lead to any significant suppression in the number of seeds compared to the baseline seeding criteria; this is likely because at these early epochs, there has not been enough build up of angular momentum in the gas to prevent seeding. As we approach lower redshifts, the suppression due to gas spin criterion becomes stronger. Additionally, we see that despite the suppression, the gas spin criterion does not change the peak epoch of seed formation i.e. $z\sim11-12$, compared to the baseline criteria; as also noted in \cite{2021arXiv210508055B}, this peak occurs because metal pollution halts the formation of new seeds at $z\lesssim11$. At $z\sim11-12$, the gas spin criterion suppresses the number of seeds by factors of $\sim6$. Lastly, the suppression is similar for all seed masses between $1.25\times10^4-8\times10^5~M_{\odot}/h$; this is due to the lack of any significant correlation between halo mass vs. gas spin seen in Section \ref{Characterizing halo properties relevant for DCBH formation}. 

Comparing the foregoing results to previous work, \cite{2006MNRAS.371.1813L} used their empirical model to predict that $\sim5\%$ of halos with $\sim10^7~M_{\odot}$ have low enough spins to form $\sim10^5M_{\odot}$ seed BHs~(this percentage increases with halo mass). In our model, an overall suppression by factors of $\sim6$ implies that about 16\% of halos satisfying the baseline criteria will actually be seeded with BHs once the gas spin criterion is applied. However, the threshold halo masses in our baseline model ($>3\times10^8~M_{\odot}/h$ for $\sim10^5~M_{\odot}$) are significantly higher than that in \cite{2006MNRAS.371.1813L} ($10^7~M_{\odot}/h$). If we reduce the halo mass threshold to $10^7M_{\odot}$, $\sim6\%$ of our halos satisfy the gas spin criterion, in good agreement with this previous work.

Next, we look at the suppression of seeding caused by the LW flux criterion~(see Figure \ref{impact_of_seeding_fig}: lower panels). Similar to the redshift trend seen with the gas spin criterion, seeding is more heavily suppressed by a lack of sufficient LW flux 
at $z\sim7-15$ compared to $z\gtrsim20$, despite the fact that LW radiation sources are more prevalent at lower redshifts. As noted in Section \ref{Characterizing halo properties relevant for DCBH formation}, this is driven by the 
reduction in LW flux~(at fixed halo mass) with decreasing redshifts due to Hubble expansion. For the same reason, the LW flux criterion pushes the peak epoch of seed formation to higher redshifts compared to the baseline criterion~(unlike the gas spin criterion). For $J_{\mathrm{crit}}=50~J_{21}$ and $100~J_{21}$, the majority of the seeds are formed around $z\sim15$ and $z\sim19$, respectively. Therefore, 
a high LW flux criterion becomes a limiting factor for seed formation 
earlier than metal enrichment. We also find that the LW flux criterion has a larger impact at lower halo mass thresholds (corresponding to lower seed masses). This is most noticeable for $J_{\mathrm{crit}}=100~J_{21}$ and is a consequence of the positive correlation between halo mass and LW flux.

We now quantify the impact of LW flux criterion by comparing it to the baseline model. For $J_{\mathrm{crit}}=50~\&~100~J_{21}$, the suppression is by factors of $\sim40$ and $\sim300$ respectively for halo mass thresholds corresponding to $M_{\mathrm{seed}}=1.25\times10^4~M_{\odot}/h$.
For halo mass thresholds corresponding to $M_{\mathrm{seed}}=1\times10^5~M_{\odot}/h$, seeds are suppressed by $\sim20$ and $\sim100$ for $J_{\mathrm{crit}}=50~\&~100~J_{21}$ respectively.   
For even higher halo mass thresholds corresponding to  $M_{\mathrm{seed}}=8 \times10^5~M_{\odot}/h$, there are no seeds formed for $J_{\mathrm{crit}}=50~\&~100~J_{21}$. For these highest seed masses, by the time halos are able to accumulate a dense, LW illuminated gas mass  of $5\times M_{\mathrm{seed}}$, they have already become significantly metal-enriched.

We can compare the results on the impact of $J_{\mathrm{crit}}$ to previous literature. When $J_{\mathrm{crit}}$ is increased from $10~J_{21}$ to $100~J_{21}$~(blue vs green lines in lower panels of Figure \ref{impact_of_seeding_fig}), the number of seeds is suppressed by factors up to $\sim100$ for $1.25\times10^4~M_{\odot}/h$ seeds, and by factors up to $\sim80$ for $1\times10^5~M_{\odot}/h$ seeds~(note however that statistical uncertainties are large for $J_{\mathrm{crit}}=100~J_{21}$). We compare this to predictions from hydrodynamic simulations of H16; notably, they are able to probe somewhat larger values of $J_{\mathrm{crit}}$~($30$ to $300~J_{21}$) due to their larger volume~($142~\mathrm{Mpc}/h$ box size). H16 find a $\sim100$ times decrease in the number densities of halos with critical LW fluxes varying from $30$ to $300$~$J_{21}$. This is broadly consistent with our results, though we note that their resolution is significantly lower than ours for the $(142~\mathrm{Mpc}/h)^3$ box. 
Our simulations are similar in resolution to those of \cite{2018ApJ...861...39D}, but we predict a stronger impact of $J_{\mathrm{crit}}$ compared to their results. More specifically, they find that the number of seeds is only suppressed by factors of $\sim7$ when  $J_{\mathrm{crit}}$ is increased from $30$ to $300$~$J_{21}$. There are some differences between our modelling and theirs that 
could potentially explain this. First, their seeding criteria 
are based on the local properties of individual gas cells; therefore, they can form seeds even if one gas element satisfies the density, metallicity and LW flux criteria. In contrast, our models require 
that a minimum total mass of gas cells~(amounting to a mass of $5~M_{\mathrm{seed}}$)  {\em simultaneously} satisfies the density, metallicity, and LW flux criteria. Second, their models allow for multiple BH seeds to form in the same halo at a given time instant, whereas our model only allows one seed per halo. Overall, these could lead to significantly fewer 
seeds formed in our model compared to \cite{2018ApJ...861...39D}, particularly for higher values of $J_{\mathrm{crit}}$. Semi-analytic models~\citep{2012MNRAS.425.2854A,2014MNRAS.443..648A,2014MNRAS.442.2036D}, on the other hand, exhibit a much stronger impact compared to our simulations as well as H16 and \cite{2018ApJ...861...39D}, with factors of $\sim10^4$ decrease in the number density of DCBH forming halos when LW flux is increased from 30 to 300 $J_{21}$~(see Figure 4 of H16). As demonstrated in H16, the differences in predictions between hydrodynamic simulations and semi-analytic models may be attributed to differences in the modelling of star formation, metal enrichment, and LW radiation. Despite these differences, all the models~(including this work) commonly predict a strong impact of LW radiation on black hole seeding. 

The impact of the LW flux criterion versus the gas spin criterion on BH seeding can be summarized as follows. First, the LW flux criterion is overall substantially more restrictive than the gas spin criterion. Second, the gas spin criterion does not impact the peak epoch of seed formation, but the LW flux criterion pushes the peak epoch of seed formation to higher redshifts. Third, the LW flux criterion preferentially suppresses seeding 
in lower mass halos~(at fixed redshift), whereas the impact of the gas spin criterion is broadly similar for all halo masses. This is primarily because halo mass does not have a significant correlation with gas spin, but it has a positive correlation with LW flux. 

Due to the lack of correlation between halo mass and gas spin, the gas spin and LW flux criteria tend 
to impact seeding independently of each other. As an example, the gas spin criterion suppresses seeding by factors of $\sim 6$ {\em regardless} of whether a LW flux criterion is applied. This can be seen by comparing the solid and dashed lines in the upper panels of Figure \ref{impact_of_seeding_fig}. Therefore, when both the gas spin and LW flux criteria are applied and compared against the baseline model, 
the suppression of seeds is a simple product of the contributions from each of two criteria, which amounts to factors of $\sim240$ and $\sim120$ for seed masses of $1.25\times10^4$ and $1\times10^5~M_{\odot}/h$, 
respectively.
\subsubsection{Seeding at higher resolution zooms}
\label{Seeding at higher resolution zooms}
\begin{figure}
\includegraphics[width=8cm]{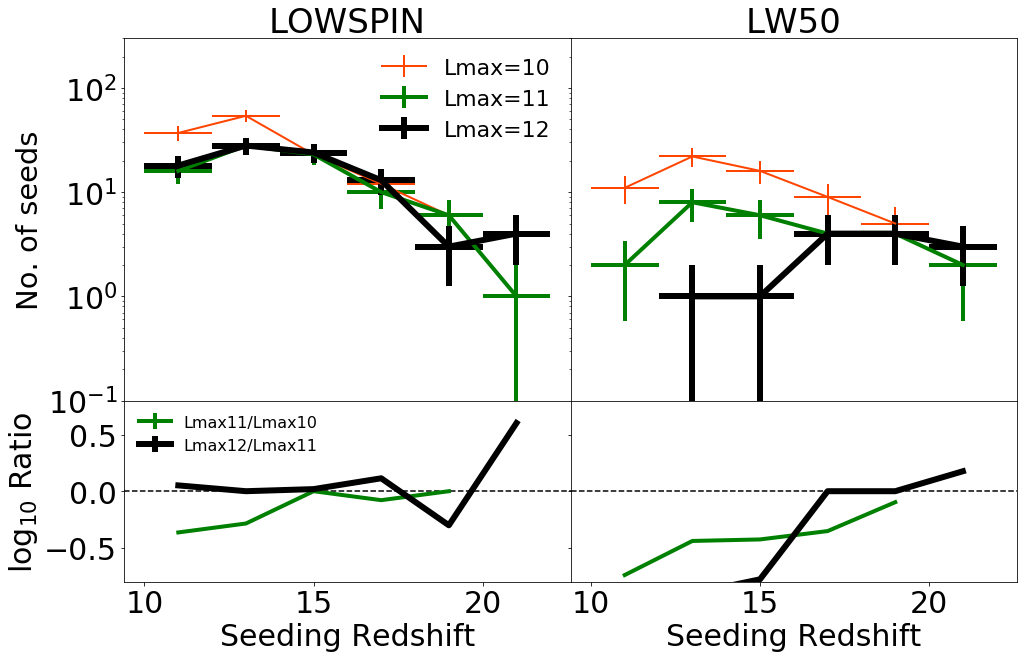} 
\caption{Resolution convergence of the distribution of seeding times for $M_{\mathrm{seed}}=10^5~M_{\odot}/h, \tilde{M}_h=3000~\&~ \tilde{M}_{\mathrm{sf,mp}}=5$. Red, green and black lines in the upper panels correspond to $L_{\mathrm{max}}=10,11~\&~12$, respectively. The left panel corresponds to the baseline + gas spin criterion; the right panel corresponds to baseline + Lyman Werner flux criterion for $J_{\mathrm{crit}}=50~J_{21}$. Green and black lines in the lower panels show the ratios between $L_{\mathrm{max}}=11$ vs. $10$ and $L_{\mathrm{max}}=12$ vs. $11$ respectively. When the gas spin criterion is applied, the simulations are reasonably well converged at $L_{\mathrm{max}}\geq11$. When the LW flux criterion is applied, there is reasonable convergence at $z\sim18-20$. However, at $z\lesssim17$ the seeding is significantly more suppressed at $L_{\mathrm{max}}=12$ compared to $L_{\mathrm{max}}=11$; this is due to relatively stronger metal enrichment at $z\lesssim17$ for $L_{\mathrm{max}}=12$, which was also seen in Figure 19 of \protect\cite{2021arXiv210508055B}.}
\label{resolution convergence seeding}
\end{figure}

We have thus far largely focused on $L_{\mathrm{max}}=11$ simulations. In \cite{2021arXiv210508055B}, we had shown that that the $L_{\mathrm{max}}=11$ results are well converged for the baseline seeding model. However, we had also seen that making the seeding criteria more restrictive~(for e.g. increasing $\tilde{M}_{\mathrm{sf,mp}}$) can reduce the rate of convergence. It is therefore instructive to also look at how the gas spin and LW flux criteria impact our resolution convergence; this is shown in Figure \ref{resolution convergence seeding} for $M_{\mathrm{seed}}=10^5~M_{\odot}/h$. We first note that adding the gas spin criteria does not significantly impact the resolution convergence; the results are convergent to within factors of $\sim1.5$. However, when a LW flux criterion is added with $J_{\mathrm{crit}}=50~J_{21}$, the resolution convergence is significantly impacted at all but the highest redshifts. At $z\sim17-20$, the $L_{\mathrm{max}}=11~\&~12$ results for the number of seeds are reasonably well converged; however, at $z\lesssim17$, seeds are much more strongly suppressed for $L_{\mathrm{max}}=12$ compared to  $L_{\mathrm{max}}=11$.

We also look at the resolution convergence of the LW flux distributions in Appendix \ref{appendix_resolution_convergence}. There, we find that the LW fluxes converge significantly more slowly within dense, metal poor gas cells, as compared to a general gas cell~(particularly for $\gtrsim50~J_{21}$). 
Nevertheless, we still find that the LW fluxes do approach convergence; therefore, we expect the seeding rates to continue converging for even higher resolutions~(albeit slowly compared to the baseline seed model). 

To explain the slower convergence rates of models with a LW flux criterion at $z\lesssim17$, we 
recall that \cite{2021arXiv210508055B}~(see Figure 19) 
found that the resolution convergence of metal enrichment at $z\lesssim17$ is slower than that of star formation. More specifically, they 
had found that while the total amount dense gas mass is well converged to within $\sim20\%$, the total dense, metal poor gas mass was $\sim2-3$ times smaller in $L_{\mathrm{max}}=12$ compared to  $L_{\mathrm{max}}=11$. Due to the faster metal enrichment in $L_{\mathrm{max}}=12$, a significantly larger fraction of LW illuminated gas cells~($\gtrsim50~J_{21}$) become metal enriched in $L_{\mathrm{max}}=12$, compared to $L_{\mathrm{max}}=11$. Therefore, applying a LW flux criterion with $J_{\mathrm{crit}}\gtrsim50~J_{21}$ tends to push seed formation to occur in regions which are metal poor at $L_{\mathrm{max}}=11$, but metal enriched at $L_{\mathrm{max}}=12$. This overall leads to a significant slow-down of resolution convergence. Pushing to higher resolutions would require a tremendous amount of computing time, memory and storage. Therefore, we continue exploring 
the trends in BH seeding within the LW flux criterion for $L_{\mathrm{max}}=11$ simulations, but we carefully account for the resolution dependence of our results when drawing conclusions. As we shall see, our main conclusions drawn from the $L_{\mathrm{max}}=11$ runs remain unchanged for $L_{\mathrm{max}}=12$.

\subsection{Varying SMBH seed masses}

\begin{figure}
 \includegraphics[width=8.5cm]{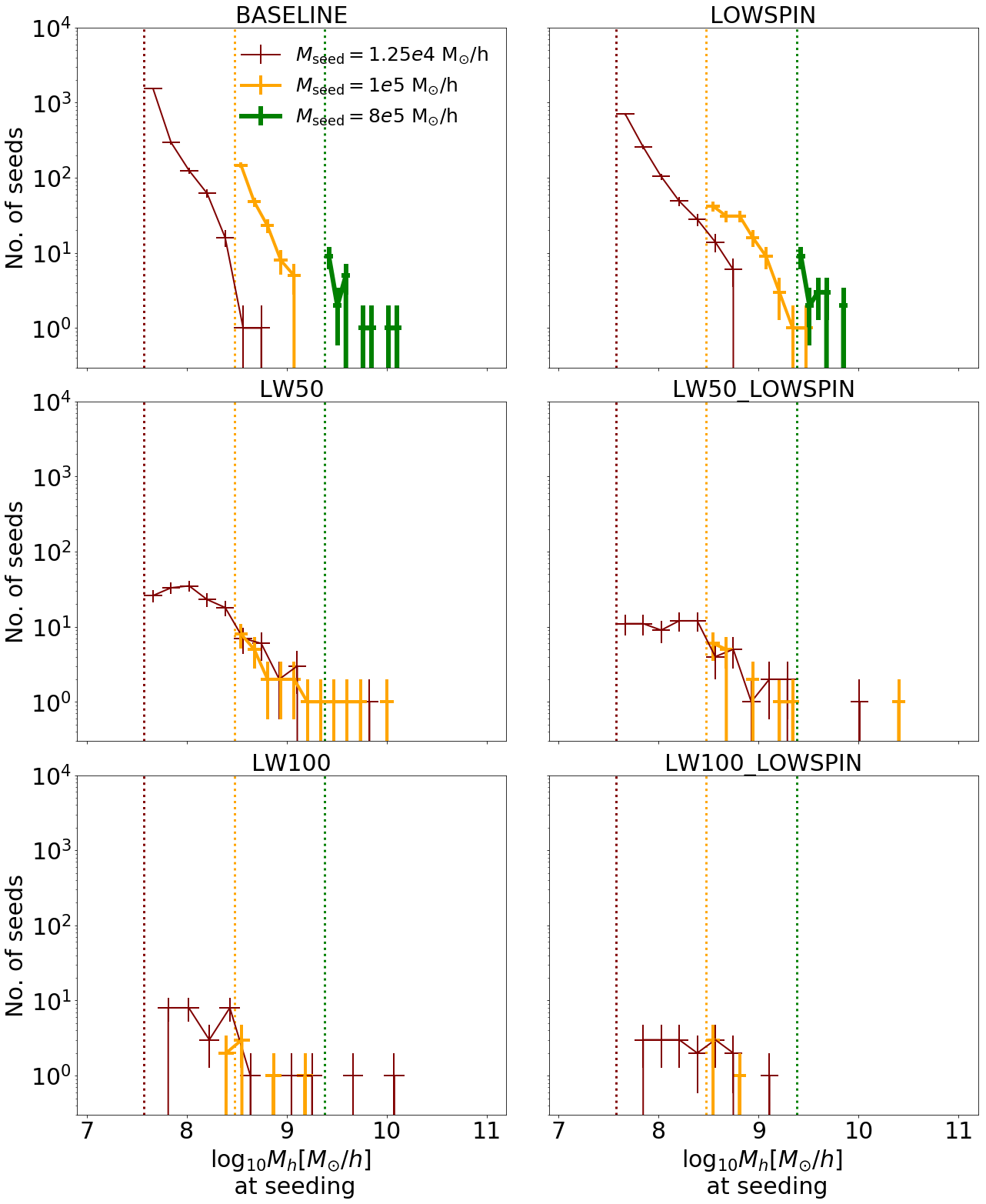} 
 \caption{Number of seeding events in various bins of host halo masses for models with $M_{\mathrm{seed}}=1.25\times10^4,~1\times10^5,~8\times10^5~M_{\odot}/h$. The vertical dotted lines show the minimum halo mass for seeding~($\tilde{M}_h\times M_{\mathrm{seed}}$). In the upper left panel, we apply only the baseline criteria for halo mass~($\tilde{M}_{h}=3000$) and star forming, metal poor gas mass~($\tilde{M}_{\mathrm{sf,mp}}=5$). In the upper right panel, we additionally apply the gas spin criterion. In the middle \& lower left panels, we additionally apply the LW flux criterion with $J_{\mathrm{crit}}=50~\&~100~J_{21}$ respectively. Lastly, the middle \& lower right panels apply both the gas spin and LW flux criteria with $J_{\mathrm{crit}}=50~\&~100~J_{21}$ respectively. When only the baseline criteria and gas spin criterion are applied, the distributions are very steep, and the majority of seeds form in halos close to the minimum mass threshold. When the LW flux criterion is added, the distributions become significantly more flat--- i.e.,  the seeding is strongly suppressed in lower mass halos~($\lesssim5\times10^8~M_{\odot}/h$) and enhanced in higher mass halos~($\gtrsim5\times10^8~M_{\odot}/h$) compared to the baseline criterion. Due to the positive correlation between LW flux and halo mass, adding a LW flux criterion pushes seed formation to happen in more massive halos. As a result, formation of the lowest-mass ($1.25\times10^4~M_{\odot}/h$) seeds is more strongly suppressed than $1\times10^5~M_{\odot}/h$ seeds.}
\label{Seed_distributions_seed_dependence_fig}
\end{figure}

\begin{figure}
\includegraphics[width=8.5cm]{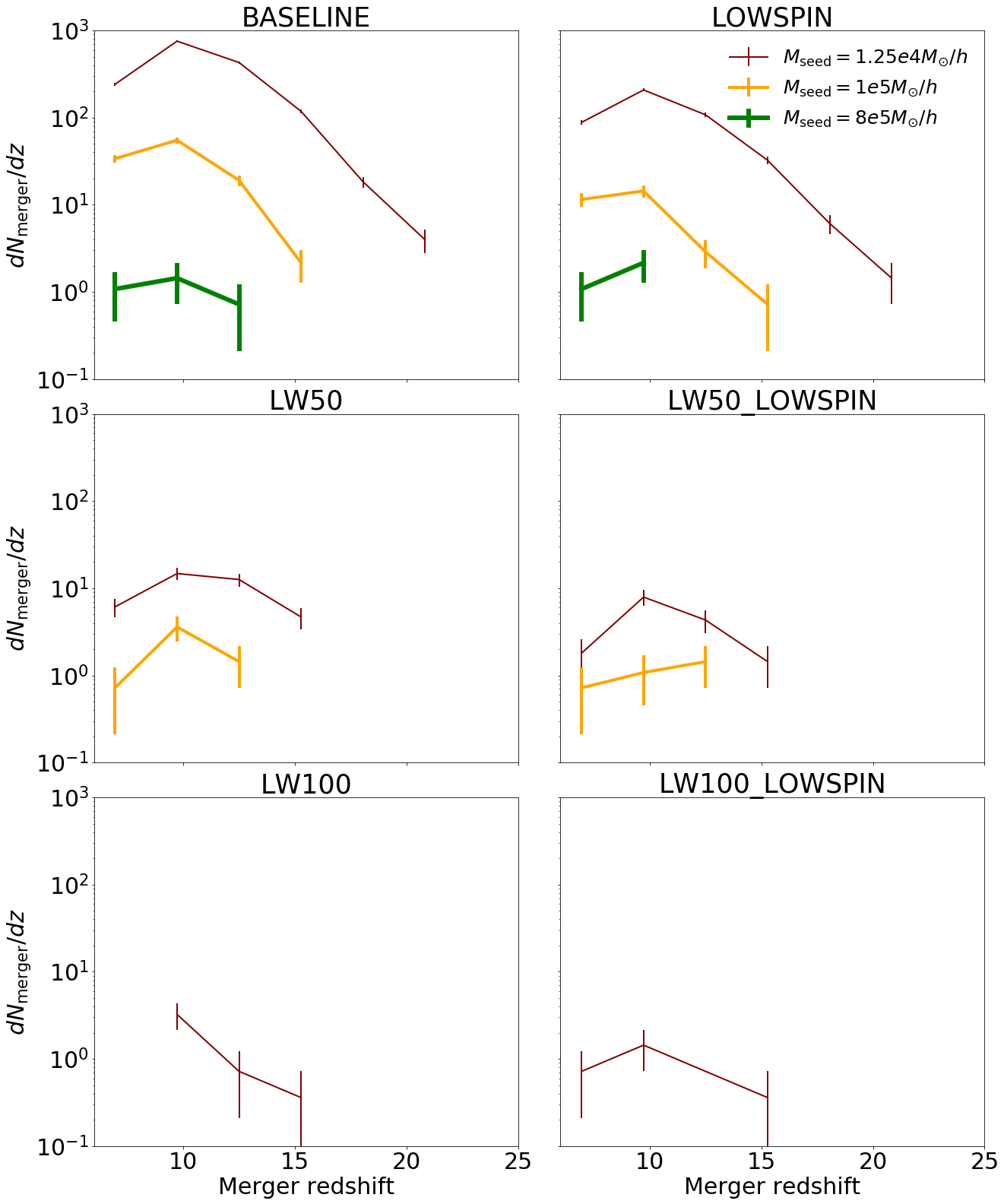} 
\caption{Comparison between BH merger rates in different seeding models, in the same format as Figure \ref{Seed_distributions_seed_dependence_fig}. 
When the gas spin criterion is applied, merger rates are overall suppressed by factors $\sim6$, and $1.25\times10^4~M_{\odot}/h$ seeds merge $\sim10$~($\sim100$) times more frequently than $1\times10^5~M_{\odot}/h$~($8\times10^5~M_{\odot}/h$) seeds. When a LW flux criterion with $J_{\mathrm{crit}}=50~J_{21}$ is applied, merger rates are overall suppressed by factors of $\sim60-100$. Because lower-mass seed formation is preferentially suppressed by the LW flux criterion, the merger rates for $1.25\times10^4~M_{\odot}/h$ seeds are still higher than the merger rates for larger seeds, but only by factors of $\sim 4$. When $J_{\mathrm{crit}}=100~J_{21}$, it leads to only a handful of mergers for $1.25\times10^4~M_{\odot}/h$ seeds, and no mergers for $1\times10^5~M_{\odot}/h$ seeds.}
\label{Merger_rates_seed_dependence_fig}
\end{figure}

\begin{figure*}

\begin{tabular}{cc}
\includegraphics[width=14cm]{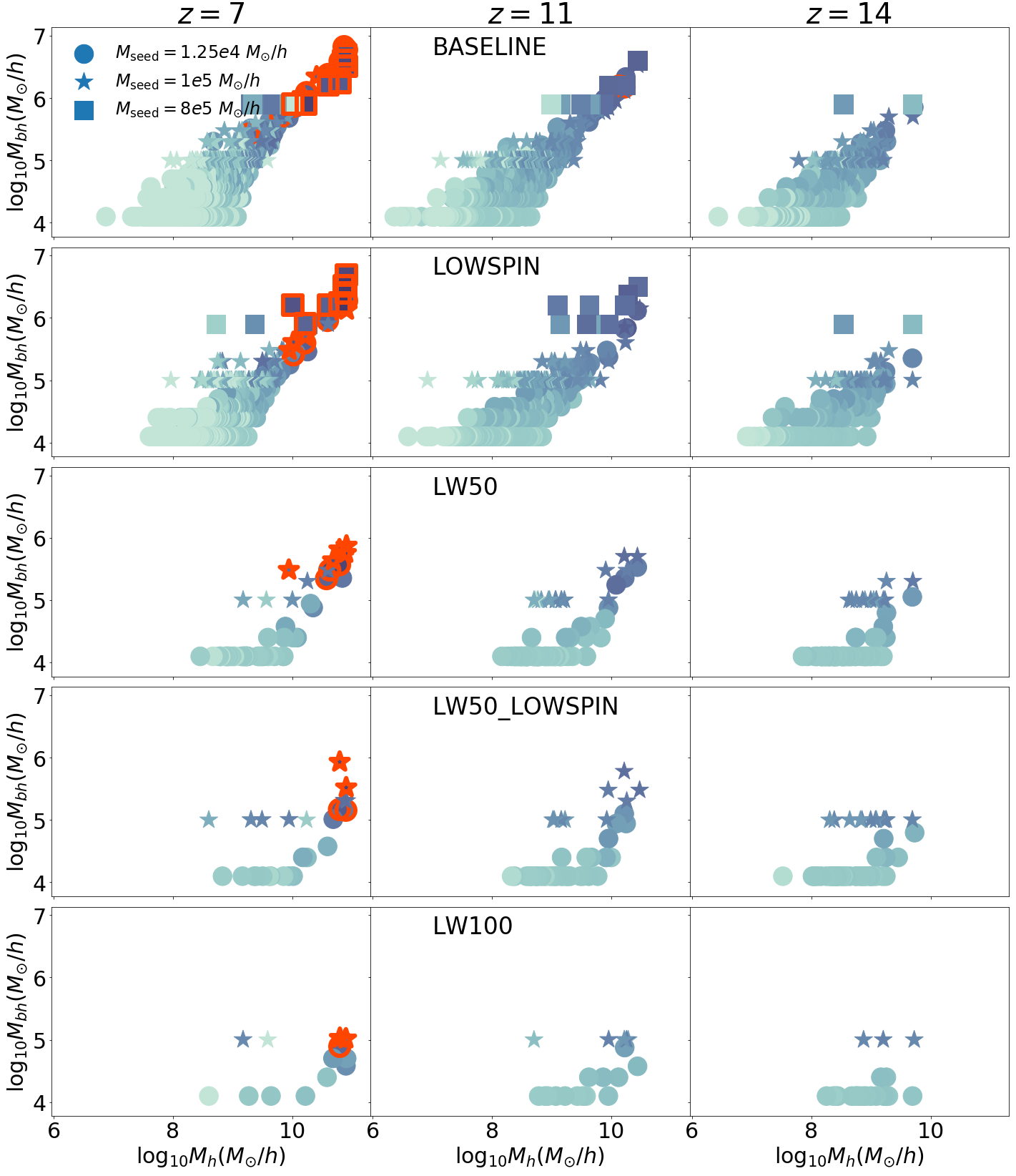} & \includegraphics[width=1.9cm]{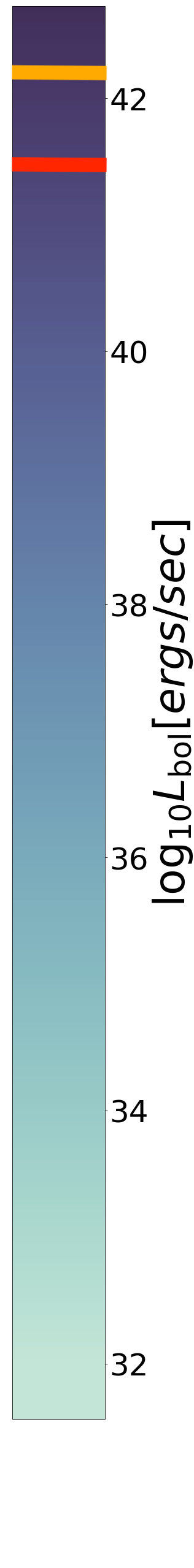} 
\end{tabular}
\caption{Halo mass vs total black hole mass relation for $M_{\mathrm{seed}}=1.25\times10^4,~1\times10^5,~8\times10^5~M_{\odot}/h$. The data points are color coded by the black hole luminosity. Only those halos are shown where $<1\%$ of the total mass is contaminated by low resolution DM particles. Left to right panels correspond to different redshift snapshots. In the 1st row, we only apply the baseline criteria for halo mass~($\tilde{M}_{h}=3000$) and star forming, metal poor gas mass~($\tilde{M}_{\mathrm{sf,mp}}=5$). In the 2nd row, we additionally apply the gas spin criterion. In the 3rd row, we apply the LW flux criterion with $J_{\mathrm{crit}}=50~J_{21}$, and we include both the LW flux and the gas spin criteria in the 4th row. In the 5th row, we apply the LW flux criterion with $J_{\mathrm{crit}}=100~J_{21}$. The red and orange  markers on the color bar correspond to the detection limit of Lynx at $z=7~\&~11$ respectively; this is assumed to be $1\times10^{-19}~\mathrm{ergs~cm^{-2}~s^{-1}}$ in the $2-10~\mathrm{keV}$ band for a survey area of $360~\mathrm{arcmin}^2$~\protect\citep{2020MNRAS.492.2535G}. The required bolometric correction is adopted from \protect\cite{2007MNRAS.381.1235V}. The points that are highlighted in red correspond to objects which surpass the Lynx detection limit. When the baseline and gas spin criteria are applied, BHs grow to $10^7~M_{\odot}/h$ for all the seed masses. When the LW flux criterion is added, $8\times10^5~M_{\odot}/h$ are absent, and the $1.25\times10^4,~1\times10^5~M_{\odot}/h$ seeds cannot grow to the supermassive regime~($\gtrsim10^6~M_{\odot}/h$) due to the absence of mergers.}%
\label{Luminosity_functions_seedmass_dependence_fig}
\end{figure*}

We finally look at the implications of the foregoing results on the predictions of merger rates, BH masses, and luminosities of $z>7$ BHs at different seed masses~(forming in halos with different total masses and dense, metal poor gas masses). 

Figure \ref{Seed_distributions_seed_dependence_fig} shows the number of seeds formed in halos of different masses for $1.25\times10^{4},1\times10^{5}~\&~8\times10^{5}~M_{\odot}/h$ seeds. When only the baseline criteria~(upper left panel of Figure \ref{Seed_distributions_seed_dependence_fig}) are applied, the distributions are very steep, and the vast majority of the seeds are forming very close to our selected halo mass threshold~($\tilde{M}_h=3000$). This continues to be true even when the gas spin criterion is added~(upper right panel of Figure \ref{Seed_distributions_seed_dependence_fig}), due to the weak correlation between halo mass and gas spin.  However, when the LW flux criterion is added with $J_{\mathrm{crit}}/J_{21}=50~\&~100$~(lower and middle panels of Figure \ref{Seed_distributions_seed_dependence_fig}), we can clearly see that the slopes of the distributions become significantly flatter. In other words,  seed formation is enhanced in higher mass halos and suppressed in lower mass halos. This is because the correlation between halo mass and LW flux requires halos to accumulate a higher mass before seeding a BH. This essentially explains why the $1.25\times10^{4}~M_{\odot}/h$ seeds have a somewhat stronger suppression than $1\times10^{5}~M_{\odot}/h$, when compared to the baseline seed model. As a result, the relative excess of $1.25\times10^{4}~M_{\odot}/h$ seeds compared to $1\times10^{5}~M_{\odot}/h$ seeds is only by factors of $\sim5$ for $J_{\mathrm{crit}}/J_{21}=50~\&~100$, in contrast to factors of $\sim20$ enhancement for low-mass seeds in the baseline seed model.

The above trends are reflected in the BH merger rates for different seed masses shown in Figure \ref{Merger_rates_seed_dependence_fig}. In the presence of only the baseline criteria and gas spin criterion~(upper panels of Figure \ref{Merger_rates_seed_dependence_fig}), merger rates of $1.25\times10^4~M_{\odot}/h$ seeds are $\sim10$ and $\sim100$ times higher compared to the merger rates for $1\times10^5~M_{\odot}/h$ and $8\times10^5~M_{\odot}/h$ seeds, respectively. When the LW flux criterion with $J_{\mathrm{crit}}=50~J_{21}$ is added~(middle panels of Figure \ref{Merger_rates_seed_dependence_fig}), merger rates are generally suppressed by factors of $\sim60-100$ compared to the baseline criterion. Additionally, because the LW flux criterion preferentially suppresses low-mass seed formation, we find that while $1.25\times10^4~M_{\odot}/h$ seeds still have the highest merger rates, they are only a factor of $\sim 4$ higher than those of $1\times10^5~M_{\odot}/h$ seeds. 

Further increasing $J_{\mathrm{crit}}$ to $100~J_{21}$~(lower panels of Figure \ref{Merger_rates_seed_dependence_fig}) causes the merger rates to be very low overall; there are only a handful of $z\gtrsim7$ mergers for $1.25\times10^4~M_{\odot}/h$ seeds, and no mergers amongst $1\times10^5~M_{\odot}/h$ seeds. Given that the inferred values of $J_{\mathrm{crit}}$ for DCBHs are much higher~($\gtrsim1000~J_{21}$) in the literature, our results imply that mergers of DCBHs would be rare and challenging for LISA to detect. Lastly, note that at these redshifts, mergers are the primary channel for BH growth in our models (see \citealt{2021arXiv210508055B} or more details); this is largely because the accretion rate scales as $M_{bh}^2$, which makes it difficult for low mass BHs to grow efficiently. 
Therefore, it is the merger rates that primarily determine the resulting final BH masses produced by the different seeds.

The final BH masses at $z=7,11,14$ produced by $1.25\times10^4,1\times10^5~\&~8\times10^5~M_{\odot}/h$ seeds are shown in Figure \ref{Luminosity_functions_seedmass_dependence_fig} for our models with different combinations of baseline seeding criteria, gas spin criterion, and LW flux criterion. When only the baseline seeding criteria are applied~(1st row of Figure \ref{Luminosity_functions_seedmass_dependence_fig}), we find that seed masses of $1.25\times10^4,1\times10^5~\&~8\times10^5~M_{\odot}/h$ grow via mergers to produce BH masses up to $10^{7}~M_{\odot}/h$ at $z\sim7-11$, reiterating the results from \cite{2021arXiv210508055B}. This continues to be true when the gas spin criteria are added~(2nd row of Figure \ref{Luminosity_functions_seedmass_dependence_fig}), and directly follows from the results of Figures \ref{Seed_distributions_seed_dependence_fig} and \ref{Merger_rates_seed_dependence_fig}. When the LW flux criterion is added~(3rd, 4th and 5th rows of Figure \ref{Luminosity_functions_seedmass_dependence_fig}), the merger-driven growth is suppressed so much that even for $J_{\mathrm{crit}}=50~J_{21}$, neither $1.25\times10^4~M_{\odot}/h$ nor  $1\times10^5~M_{\odot}/h$ seeds are able to form SMBHs of masses $\gtrsim10^6~M_{\odot}/h$ by $z=7$ in our simulation volume. Lastly, due to the stronger suppression of seed formation and merger rates of lower mass seeds for $J_{\mathrm{crit}}/J_{21}=50~\&~100$ we see that the lowest-mass $1.25\times10^4~M_{\odot}/h$ seeds end up producing slightly smaller final BH masses~(by factors of $\sim2-4$) compared to $1\times10^5~M_{\odot}/h$ seeds at $z\sim7-11$~(although statistics are limited). Recall that $8\times10^5~M_{\odot}/h$ seeds are completely absent for $J_{\mathrm{crit}}/J_{21}=50~\&~100$.

Finally, we look at the BH luminosities produced by $1.25\times10^4,1\times10^5~\&~8\times10^5~M_{\odot}/h$ seeds~(color coded in the data points of Figure \ref{Luminosity_functions_seedmass_dependence_fig}). These luminosities were estimated from the BH accretion rates using Eq.~(\ref{bolometric_lum_eqn}). For models with only the baseline seeding criteria and gas spin criteria, all three seed masses produce BHs reaching luminosities of up to $\sim10^{42}~\mathrm{ergs~s^{-1}}$ and $\sim10^{43}~\mathrm{ergs~s^{-1}}$ at $z=11$ and $z=7$ respectively. When the LW flux criteria with $J_{\mathrm{crit}}=50~\&~100~J_{21}$ are applied, the luminosities~(at fixed halo mass) drop by a factor of $\sim10~\&~100$ respectively compared to the baseline criterion~(due to the drop in BH masses). We also compare these luminosities to the detection limit of Lynx, which is $1\times10^{-19}~\mathrm{ergs~cm^{-2}~s^{-1}}$ in the $2-10~\mathrm{keV}$ band for a survey area of $360~\mathrm{arcmin}^2$~(marked in the color-bar of Figure \ref{Luminosity_functions_seedmass_dependence_fig} for $z=7,11$). Note that our results here are subject to theoretical uncertainties in our BH accretion model as well as the bolometric corrections, which are adopted from \cite{2007MNRAS.381.1235V}. At $z\geq11$, even for the baseline criterion, there are no BHs above the Lynx detection limit. At $z=7$ where somewhat lower luminosities can be detected, we do have detectable BHs; but in the presence of LW flux criterion with $J_{\mathrm{crit}}/J_{21}=50~\&~100$, their number reduces to only a handful. Given the much higher $J_{\mathrm{crit}}$ values of $\gtrsim1000~J_{21}$ inferred in the literature, our results suggest that Lynx will not find any detectable DCBHs within regions with overdensities similar to or lower than \texttt{ZOOM_REGION_z5}. In future work, we plan to explore the detectability of DCBHs in more extreme cosmological regions. 

The key takeaway is that even for relative low values of $J_{\mathrm{crit}}$~($\gtrsim50~J_{21})$, our simulations with LW flux criteria fail to produce BHs in the supermassive regime~($\gtrsim10^6~M_{\odot}/h$) by $z\sim7$. Seeds with $8\times10^5~M_{\odot}/h$ completely fail to form; $1.25\times10^4~M_{\odot}/h$ and $1\times10^5~M_{\odot}/h$ seeds do form but are not able to grow to the supermassive regime. The growth is further suppressed at higher resolutions, where even fewer seeds form. We again emphasize that the foregoing  results are 
specific to our underlying assumptions, including Bondi accretion, which struggles to grow low mass BHs at early times due to the $\sim M_{bh}^2$ scaling of the accretion rate. Accretion rates are decreased further due to the fact that for these early protogalaxies, the halo centers (where BHs are repositioned to) may be offset from the densest gas within the halo by distances up to $\sim10~\mathrm{kpc}/h$. Additionally, 
our conclusions do not necessarily apply to the regime of the observed $z\gtrsim7$ quasars with BH masses up to $\sim10^9~M_{\odot}/h$. This is because high-z quasars are expected to reside in much more extreme regions than our zoom volume, where accretion will have a more significant~(and potentially dominant) contribution. This means that our conclusions may change when our models are applied to these extreme regions, and do not yet rule out DCBH seeds as progenitors of $z\gtrsim7$ quasars; we are exploring this in an ongoing work. 



\section{Summary and Discussion}
\label{Summary and Conclusions}

In this work, we quantify the of impact of gas spin and LW flux based seeding conditions on the $z\geq7$ SMBH populations using cosmological hydrodynamical zoom simulations, and we assess the implications of our results for DCBH seed formation.

The zoom region was selected to produce a target halo of mass $3.5\times10^{11}~M_{\odot}/h$ at $z=5$~(corresponding to a peak height of $\nu=3.3$). We then start with a set of baseline seeding criteria as outlined in \cite{2021arXiv210508055B} to ensure that seeds are formed only in pristine halos with \textit{dense gas} i.e exceeding the star formation threshold of $0.1~\mathrm{cm}^{-3}$: BH seeding sites are required to have a minimum threshold of total halo mass~($3000~M_{\mathrm{seed}}$) and dense, metal poor gas mass~($5~M_{\mathrm{seed}}$). The baseline criteria enforces seeding to take place only in halos that have grown significantly~(by factors of $\gtrsim4-10$ depending on the seed mass) since crossing the atomic cooling threshold~($\sim10^7~M_{\odot}/h$). These halos have a prior history of star formation and metal enrichment. However, metals fail to pollute the entire halo, leaving behind pockets of dense, metal poor gas embedded within star forming regions. These dense, metal poor, LW illuminated pockets have gas masses ranging from $\sim10^5-10^6~M_{\odot}/h$. 

We then add the gas spin and LW flux criteria, and focus on their impact on seed formation and the resulting $z\gtrsim7$ SMBH populations. These are described as: 
\begin{itemize}
\item \textit{Gas spin 
criterion}: The dimensionless spin angular momentum~($\lambda$) of the gas in the host halo must be less than the minimum value~($\lambda_{\mathrm{max}}$) required for the gas disc to be gravitationally stable. 

\item \textit{LW flux criterion}: The minimum threshold ($5~M_{\mathrm{seed}}$) for dense, metal poor gas mass within host halos must also be illuminated by LW intensities greater than a critical flux $J_{\mathrm{crit}}$. Star formation is suppressed within all gas cells exposed to the supercritical LW flux. 
\end{itemize}
We explored a wide range of models with the gas spin criterion and LW flux criterion~($J_{\mathrm{crit}}/J_{21}=10,50~\&~100$) using seed masses of $1.25\times10^4$, $1\times10^5$ and $8\times10^5~M_{\odot}/h$. This exploration was carried out at gas mass resolutions of $\sim10^4~M_{\odot}/h$ within our zoom region. Our key findings are as follows:

\begin{enumerate}
\item When seeding is limited to halos with low gas spin ($\lambda < \lambda_{\rm max}$), the overall rates of seed formation are suppressed by factors of $\sim6$ for all seed masses, particularly at $z\sim11-12$ when most seeds form. The suppression is similar for all seed masses / halo masses, because the correlation between halo mass and gas spin is weak. Additionally, the gas spin criterion has a weaker effect at higher redshifts and is negligible at $z\gtrsim20$.

\item The LW flux criterion has a substantially stronger impact~(compared to the gas spin criterion) on seed formation rates, even for relatively low values of $J_{\mathrm{crit}}/J_{21}$ such as $50~\&~100$; this restricts seed formation to occur only in halos~(typically $\gtrsim10^8~M_{\odot}/h$) that have enough LW sources to provide the necessary 
fluxes to halt star formation within dense, metal poor pockets of gas. For $J_{\mathrm{crit}}=50~J_{21}$, formation of $8\times10^5~M_{\odot}/h$ seeds is completely suppressed, while $1.25\times10^4~M_{\odot}/h$ and $1\times10^5~M_{\odot}/h$ seeds are suppressed by factors of $\sim40$ and $\sim20$, respectively. The formation of lower-mass seeds in lower-mass halos is preferentially suppressed, because higher mass halos have more star forming gas and are therefore more likely to provide the critical LW flux to the metal poor pockets. 


\item When both the gas-spin and LW flux criteria~($J_{\mathrm{crit}}=50~J_{21}$) are imposed (in addition to the baseline model), seed formation is even more strongly suppressed. Relative to the baseline model, seeding events are 
suppressed by factors of $\sim240$ and $\sim120$ for seed masses of $1.25\times10^4$ and $1\times10^5$ respectively.

\item Merger rates for all seed masses are suppressed by factors of $\sim6$ when seeding is only limited by the gas spin criterion and not limited by a LW flux criterion. In this case, lower-mass ($1.25\times10^4~M_{\odot}/h$) seeds merge $\sim10$~($\sim100$) times more frequently than $1\times10^5~M_{\odot}/h$~($8\times10^5~M_{\odot}/h$) seeds. In contrast, when the seeds are limited to halos with LW fluxes $J>J_{\mathrm{crit}}=50~J_{21}$, the merger rates are 
suppressed by factors of $\sim60-100$ compared to the baseline criterion.  $1.25\times10^4~M_{\odot}/h$ seeds are suppressed somewhat more strongly, but still merge $\sim4$ times more frequently than 
$1\times10^5~M_{\odot}/h$ seeds. 
With a higher $J_{\mathrm{crit}}$ value of $100~J_{21}$, there are only a handful of mergers for $1.25\times10^4~M_{\odot}/h$ seeds and none for $1\times10^5~M_{\odot}/h$ and $8\times10^5~M_{\odot}/h$ seeds.

\item When only the baseline seeding criteria and gas spin criterion are applied, all 
seed masses~($1.25\times10^4,1\times10^5~\&~8\times10^5~M_{\odot}/h$) form SMBHs up to masses of  $\sim10^7~M_{\odot}/h$ at $z\sim7-11$. %
With the addition of a LW flux criterion, due to the absence of $8\times10^5~M_{\odot}/h$ seeds and lack of mergers among $1.25\times10^4$ and $1\times10^5~M_{\odot}/h$ seeds, none of the BHs reach the supermassive regime~($\gtrsim10^6~M_{\odot}/h$) by $z\sim7$ in our simulations. 

\end{enumerate}

Our results for the gas spin criterion are reasonably well converged at the fiducial gas mass resolutions of $\sim10^{4}~M_{\odot}/h$. However, when the LW flux criterion~($J
_{\mathrm{crit}}\geq50~J_{21}$) is added, the resolution convergence is substantially slower. 
More specifically, at higher resolutions~(gas mass resolutions of  $\sim10^{3}~M_{\odot}/h$), the LW flux criterion produces an even stronger suppression of seeding at $z\lesssim17$. This is primarily driven by more rapid metal enrichment at higher resolutions. Nevertheless, we expect our results to continue to converge at even higher resolutions. Lastly we also note that despite the slower resolution convergence, the main qualitative conclusions drawn at the fiducial resolution remain unchanged at higher resolutions. These are summarized in the following paragraph.

Overall, we find that both the gas spin and LW flux criteria significantly impact BH seed formation. The LW flux criterion tends to have a much stronger impact. Even for critical fluxes as low as $50~J_{21}$, we see a complete absence of $8\times10^5~M_{\odot}/h$ seeds and a drastic suppression in $10^4-10^5~M_{\odot}/h$ seeds; as a result, no BHs grow to the supermassive~($\gtrsim10^6~M_{\odot}/h$) regime by $z\sim7$. Recall again that for realistic galaxy spectra at these redshifts, the inferred values of $J_{\mathrm{crit}}$ are much higher~($\sim1000~J_{21}$) compared to the values adopted in this work. It is clear from our results that for such high $J_{\mathrm{crit}}$, a larger or more highly biased cosmological volume would be required to model seed formation. 
Therefore, our findings agree with the general consensus that conditions for DCBH seed formation are very restrictive~\citep[e.g.,][and references therein]{doi:10.1146/annurev-astro-120419-014455}. Without a more significant contribution to BH growth from gas accretion, it would be challenging to explain a sizable majority of $z>7$ SMBHs solely using DCBH channels. 

Our findings are quantitatively consistent with some of the previous works, but have differences compared to others. For instance, the impact of gas spin predicted by our model is similar to that of \cite{2006MNRAS.371.1813L}. In terms of the LW flux criterion, we predict a similar impact as the hydrodynamic simulations of H16. Note however that due to a relatively small zoom volume, our results are not as statistically robust for the highest fluxes~($\gtrsim100~J_{21}$) compared to the largest volume of H16~($142~\mathrm{Mpc}/h$ per side). We however predict a somewhat stronger impact compared to \cite{2018ApJ...861...39D}. This may be because our seeding criteria becomes significantly more strict for higher $J_{\mathrm{crit}}$ since we require a much larger minimum mass of dense, metal poor, LW illuminated gas to insert seeds; in contrast, \cite{2018ApJ...861...39D} imposes a similar criteria but only for one individual gas particle at mass resolutions similar to ours. Semi-analytic models \citep{2012MNRAS.425.2854A,2014MNRAS.443..648A,2014MNRAS.442.2036D} predict a substantially stronger impact of LW flux compared to hydrodynamic simulations; this can be attributed to differences in the modelling of star formation, metal enrichment, and LW radiation~(as demonstrated by H16). Regardless of the quantitative differences, all of these works agree that both gas spin and LW flux can have a substantial impact on seed formation. Our work additionally demonstrates that the LW flux criterion tends to be more restrictive compared to the gas spin criterion.   

Our results have several potential implications for upcoming observational facilities. The fact that the LW flux criterion pushes seed formation to higher-mass halos may have two important observational consequences. First, higher mass halos are more rare, so the resulting merger rates are very low; this implies that LISA may find it much more challenging to detect mergers originating from DCBH channels, compared to other channels~(e.g. Pop III). Second, for the events that are detected by LISA, follow-up electromagnetic observations of their host galaxies using JWST may be useful for constraining their seeding origins. Electromagnetic observations may also be able to distinguish other signatures of the DCBH seeding models presented here.

The AGN luminosities drop by factors of $\sim10$ and $\sim100$ for $J_{\mathrm{crit}}=50~\&~100~J_{21}$ respectively compared to the baseline criterion. As a result, for $J_{\mathrm{crit}}=50~\&~100~J_{21}$, our zoom volume produces very few objects that would be detectable with Lynx. We expect the suppression in luminosities to be even higher for larger values of $J_{\mathrm{crit}}$. Larger uniform volume simulations will be required to constrain the high-redshift AGN luminosity function resulting from our seed models, which we will explore in future work. Overall, our results suggest that if DCBHs indeed form only in the presence of very high LW fluxes~($\gtrsim1000~J_{21}$), future electromagnetic observational facilities will find it challenging to detect DCBHs. 

Our results are strongly influenced by the fact that at these early epochs, BH growth is dominated by mergers, and there is very little growth due to gas accretion. This is a well-known issue in simulating the growth of low-mass BH seeds, which owes in large part to the $M_{\rm BH}^2$ scaling of the Bondi-Hoyle accretion model. A variety of alternate accretion models exist in the literature \citep{2007ApJ...665..107P, 2009MNRAS.398...53B, 2017MNRAS.470.1121T, 2020arXiv201201458Z}, but the $M_{\rm BH}^2$ scaling is generic to all models in which the gas capture radius is assumed to scale with BH mass. 

Alternate accretion models exist that can have much smaller scaling exponents for the accretion rate vs. BH mass~(e.g. $\propto M_{bh}^{1/6}$ for accretion driven by stellar gravitational torques \citep{2011MNRAS.415.1027H,2017MNRAS.464.2840A,2019MNRAS.486.2827D}). If such a model can be reliably applied in the high-redshift regime, this might significantly boost early growth. Additionally, we note that both accretion rates and merger rates are likely to be influenced by the BH repositioning scheme, which causes the BHs to rapidly sink to the halo centers. Recent work incorporating more realistic subgrid prescriptions for modeling unresolved dynamical friction has been shown to increase merger times and decrease accretion rates \citep{2017MNRAS.470.1121T,2021arXiv210209566B}. We plan to explore the impact of BH dynamics on early seed growth in future work. 

BH accretion rates can also be impacted by the modelling of BH dynamics. Recall that the BHs are repositioned to the halo center~(minimum potential). But in these early proto-galaxies, the halo center does not always co-incide with the densest gas cell. This makes the BHs wander around~(up to distances $\sim10~\mathrm{kpc}/h$) the regions with highest gas density, which further reduces the accretion rates. At the same time, recent works~\citep{2018ApJ...857L..22T,2021MNRAS.503.6098R,2021arXiv210400021C,2021arXiv211014154N,2021MNRAS.508.1973M} that implement more realistic BH dynamics models find that BHs may be offset from the halo center for a substantial amount of time, particularly for clumpy high-z galaxies. This would also lead to reduction or delays in BH mergers and slow down the merger driven BH growth. 

We again emphasize that molecular~($H_2$) cooling, which is a crucial component for DCBH formation, is not explicitly included in our model. This artificially suppresses star formation in mini-halos~($M_h\sim10^5-10^6~M_{\odot}$); since mini-halos are progenitors of atomic cooling halos~($T_{\mathrm{vir}}\gtrsim10^4~\mathrm{K}$ or $M_h\gtrsim10^7~M_{\odot}$), this could lead to an artificially higher number of metal poor halos that cross the atomic coolong threshold and potentially overestimate the number of seeds formed. However, note that our seeds are largely forming in $\sim10^8-10^{10}~M_{\odot}/h$ halos wherein a significant amount of time has passed since they crossed the atomic cooling threshold and had their first burst of star formation. Yet, we see that they do not get completely polluted with metals, which creates the opportunity for seed formation to occur within dense, metal poor pockets. For these halos, our underlying galaxy formation model~\citep{2003MNRAS.339..289S} is well calibrated to account for the delay in star formation caused at the mini-halo stage. That being said, we do note that a further reduction in the seed mass or the halo mass threshold could cause seeds to form closer to the atomic cooling threshold; in this case, the lack of star formation in mini halos would be a more serious issue. In the future, we plan to assess this in more detail in future work with galaxy formation models that do include $H_2$ cooling.

Related to the above, our seed formation scenario is somewhat distinct from what has been explored in previous works. For example, a commonly considered scenario~(``synchronised pair scenario") is that a halo with no prior star formation history forms a DCBH as soon as it crosses the atomic cooling threshold, if it receives LW radiation from a nearby star-forming halo that has also crossed 
the threshold within the last $\sim5~\mathrm{Myr}$ ~\citep{2017NatAs...1E..75R,2021MNRAS.503.5046L,2014MNRAS.445.1056V}. In contrast, our simulations probe DCBH forming conditions in halos that have grown well past the atomic cooling threshold; these halos have a prior star formation and metal enrichment history but still contain pockets of dense, metal poor gas. These pockets form seeds upon receiving LW radiation from surrounding star forming regions within the same halo. This additional scenario also indicates that DCBHs may be slightly less rare than previously thought. Future works with explicit molecular cooling recipes will enable us to also probe DCBH formation in halos close to the atomic cooling threshold via the synchronised pair scenario.

 We also emphasize that the results of this work should not be extrapolated to the regime of observed high redshift quasars, since they are likely a tiny fraction of the overall SMBH population forming in regions much more overdense than our zoom volume. In such extreme regions, we can expect gas accretion to have an increasingly significant~(and potentially dominant) role in the BH growth. In future work, we plan to explore more extreme overdense regions, which could probe much higher $J_{\mathrm{crit}}$ values~($\gtrsim1000~J_{21}$) that are representative of actual DCBH formation conditions based on radiation hydrodynamic simulations and one-zone chemistry models~\citep{2010MNRAS.402.1249S,2014MNRAS.445..544S,2017MNRAS.469.3329W}.

Lastly, while this work is largely motivated by the DCBH seeding channel, it is part of a continued series of studies on the underlying seeding prescriptions, agnostic about the physical channels they may represent. Between this work and \cite{2021arXiv210508055B}, we have now expanded our seeding models to encompass most of the physical properties commonly associated with theoretical gas-dependent BH seed formation channels. These works will serve as a basis for continued development of seeding prescriptions, particularly in the context of large volume uniform simulations.

\begin{figure*}
\includegraphics[width=16cm]{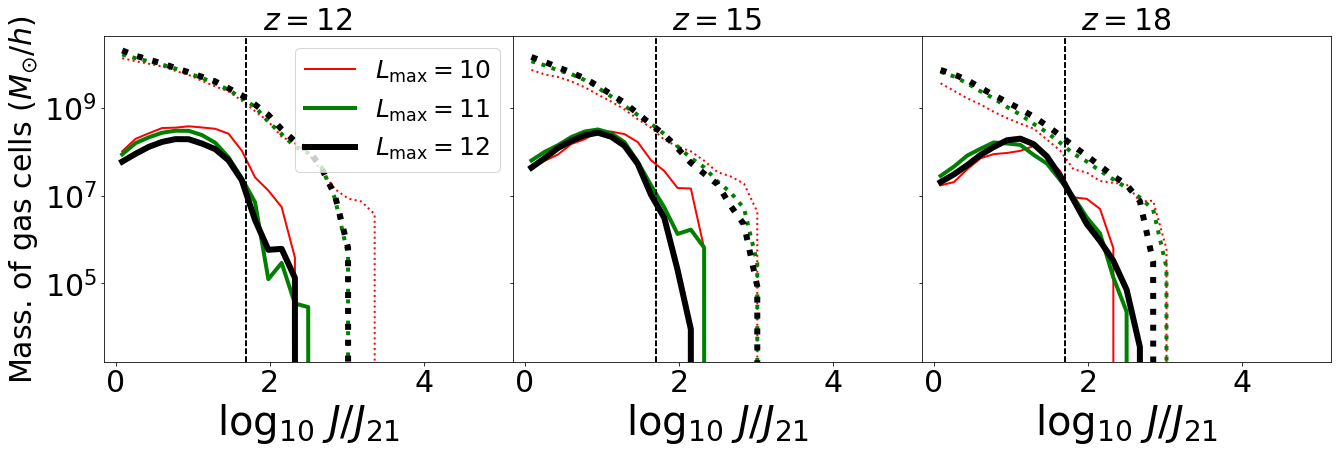} 
\caption{Resolution convergence of the distribution of LW flux values amongst all gas cells~(dashed lines) and dense, metal poor gas cells~(solid lines). Red, green and black lines in the upper panels correspond to $L_{\mathrm{max}}=10,11~\&~12$ respectively. The flux distributions among all gas cells~(dashed lines) are reasonably well converged between $L_{\mathrm{max}}=11$ and $12$. The flux distributions among dense, metal poor gas cells do converge, but at substantially slower rate for $\gtrsim50~J_{21}$. Since BHs are seeded based on LW fluxes within dense, metal poor gas cells, resolution convergence of seeding rates at $z\lesssim17$ is significantly slower.}
\label{resolution convergence}
\end{figure*}

\section*{Acknowledgements}
LB~acknowledges support from National Science Foundation grant AST-1715413. LB and PT acknowledges support from NSF grant AST-1909933 and NASA ATP Grant 80NSSC20K0502. PT also acknowledges support from AST-200849. DN acknowledges funding from the Deutsche Forschungsgemeinschaft (DFG) through an Emmy Noether Research Group (grant number NE 2441/1-1). MV acknowledges support through NASA ATP grants 16-ATP16-0167, 19-ATP19-0019, 19-ATP19-0020, 19-ATP19-0167, and NSF grants AST-1814053, AST-1814259,  AST-1909831 and AST-2007355.

\section*{Data availablity}
The underlying data used in this work shall be made available upon reasonable request to the corresponding author.

\bibliography{references}

\appendix

\section{Resolution convergence of LW intensity calculation}
\label{appendix_resolution_convergence}
In Section \ref{Seeding at higher resolution zooms}, we found that the resolution convergence of the BH seeding rates between $L_{\mathrm{max}}=11~\&~12$ is significantly worse at $z\gtrsim17$ in the presence of a LW flux criterion with $J_{\mathrm{crit}}=50~J_{21}$. Given that the baseline models of \cite{2021arXiv210508055B} were reasonably well converged for $L_{\mathrm{max}}\geq11$, we find it instructive to look at the resolution convergence of our calculated LW fluxes. Figure \ref{resolution convergence} shows the distributions of LW flux values amongst all gas cells~(dotted lines) and dense, metal poor gas cells~(solid lines). The LW flux distributions among all gas cells are reasonably well converged between $L_{\mathrm{max}}=11$ and $12$. But the convergence of seeding rates depends only on the LW fluxes among dense, metal poor gas cells. The LW flux distributions among dense, metal poor gas cells do converge, but significantly more slowly than than the LW flux distributions among all gas cells~(solid lines vs dashed lines in Figure \ref{resolution convergence}); this is particularly true for LW fluxes $\gtrsim50~J_{21}$.

The reason for the slower resolution convergence of LW fluxes among dense, metal poor gas cells is that 
metal enrichment at $z\lesssim17$ occurs faster at $L_{\mathrm{max}}=12$ compared to $L_{\mathrm{max}}=11$. As a result, a significant fraction of the gas that is metal-poor and LW-irradiated in $L_{\mathrm{max}}=11$ simulations has, at the same epoch, already become 
 metal enriched in the $L_{\mathrm{max}}=12$ simulations. Overall, this explains why resolution convergence between $L_{\mathrm{max}}=11$ and $12$ of BH seeding is substantially slow in the presence of a LW flux criterion with $J_{\mathrm{crit}}=50~J_{21}$; nevertheless, the convergence of the LW fluxes in Figure \ref{resolution convergence} hints that we can expect the seeding rates to continue converging at even higher resolutions~($L_{\mathrm{max}}\geq13$).

\end{document}